\documentclass[twocolumn,showpacs,preprintnumbers,amsmath,amssymb,pra,aps,superscriptaddress]{revtex4-1}

\usepackage{graphicx}
\usepackage{amsmath}
\usepackage{verbatim}
\usepackage{bm}
\usepackage{siunitx}
\usepackage[space]{grffile}
\usepackage{xr} 
\usepackage{color}
\usepackage{hyperref}
\usepackage{cleveref}

\graphicspath{{../figs/}}

\newcommand{\ket}[1]{|#1\rangle}
\newcommand{\bra}[1]{\langle#1|}
\newcommand{\XX}{\langle X\texttt{$\otimes$} X\rangle}
\newcommand{\YY}{\langle Y\texttt{$\otimes$} Y\rangle}
\newcommand{\ZZ}{\langle Z\texttt{$\otimes$} Z\rangle}
\newcommand{\SXX}{\mathrm{XX}}
\newcommand{\SZZ}{\mathrm{ZZ}}
\newcommand{\HZZ}{H_{\SZZ}}
\newcommand{\HZZXX}{H_{\SZZ,\SXX}}
\newcommand{\sD}{s_{\mathrm{D}}}
\newcommand{\FBell}{F_{\mathrm{\ket{\Phi^+}}}}
\newcommand{\Lcomp}{L_{\mathrm{comp, Q}}}
\newcommand{\Lcompth}{L^{\mathrm{th}}_{\mathrm{comp,Q}}}
\newcommand{\Lplus}{\ket{\Phi^+}}

\newcommand{\MA}{M_{\mathrm{A}}}
\newcommand{\QDH}{Q_{\mathrm{DH}}}
\newcommand{\QDL}{Q_{\mathrm{DL}}}
\newcommand{\QA}{Q_{\mathrm{A}}}
\newcommand{\vecs}{\vec{M}_A}
\newcommand{\TPR}{\mathrm{TPR}}
\newcommand{\FPR}{\mathrm{FPR}}
\newcommand{\fpost}{f_{\mathrm{post}}}
\newcommand{\numstates}{N_h}
\newcommand{\numoutputs}{N_o}
\newcommand{\pleak}{p_{\mathrm{leak}}}
\newcommand{\pseep}{p_{\mathrm{seep}}}

\newcommand{\nA}{n_{\mathrm{A}}}

\crefname{equation}{Eq.}{Eqs.}
\crefname{figure}{Fig.}{Figs.}
\crefname{tabular}{Tab.}{Tabs.}
\crefname{table}{Table}{Tables}
\crefname{section}{Sec.}{Secs.}

\renewcommand{\thetable}{\arabic{table}}

\begin{document}

\title{Protecting quantum entanglement from leakage and qubit errors via repetitive parity measurements}
\author{C.~C.~Bultink}
\affiliation{QuTech, Delft University of Technology, P.O. Box 5046, 2600 GA Delft, The Netherlands}
\affiliation{Kavli Institute of Nanoscience, Delft University of Technology, P.O. Box 5046, 2600 GA Delft, The Netherlands}
\author{T.~E.~O'Brien}
\affiliation{Instituut-Lorentz for Theoretical Physics, Leiden University, P.O. Box 9506, 2300 RA Leiden, The Netherlands}
\author{R.~Vollmer}
\author{N.~Muthusubramanian}
\author{M.~W.~Beekman}
\author{M.~A.~Rol}
\affiliation{QuTech, Delft University of Technology, P.O. Box 5046, 2600 GA Delft, The Netherlands}
\affiliation{Kavli Institute of Nanoscience, Delft University of Technology, P.O. Box 5046, 2600 GA Delft, The Netherlands}
\author{X.~Fu}
\affiliation{QuTech, Delft University of Technology, P.O. Box 5046, 2600 GA Delft, The Netherlands}
\author{B.~Tarasinski}
\author{V.~Ostroukh}
\author{B.~Varbanov}
\author{A.~Bruno}
\author{L.~DiCarlo}
\affiliation{QuTech, Delft University of Technology, P.O. Box 5046, 2600 GA Delft, The Netherlands}
\affiliation{Kavli Institute of Nanoscience, Delft University of Technology, P.O. Box 5046, 2600 GA Delft, The Netherlands}

\date{\today}

\begin{abstract}
Protecting quantum information from errors is essential for large-scale quantum computation.
Quantum error correction (QEC) encodes information in entangled states of many qubits, and performs parity measurements to identify errors without destroying the encoded information.
However, traditional QEC cannot handle leakage from the qubit computational space. 
Leakage affects leading experimental platforms, based on trapped ions and superconducting circuits, 
which use effective qubits within many-level physical systems.
We investigate how two-transmon entangled states evolve under repeated parity measurements, and demonstrate the use of hidden Markov models to detect leakage using only the record of parity measurement outcomes required for QEC.
We show the stabilization of Bell states over up to 26 parity measurements by mitigating leakage using postselection, and correcting qubit errors using Pauli-frame transformations. 
Our leakage identification method is computationally efficient and thus compatible with real-time leakage tracking and correction in larger quantum processors.
\end{abstract}

\maketitle

\section{Introduction}
Large-scale quantum information processing hinges on overcoming errors from environmental noise and imperfect quantum operations.
Fortunately, the theory of QEC predicts that the coherence of single degrees of freedom (logical qubits) can be better preserved by encoding them in ever-larger quantum systems (Hilbert spaces), provided the error rate of the constituent elements lies below a fault-tolerance threshold~\cite{Terhal15}. 
Experimental platforms based on trapped ions and superconducting circuits have achieved error rates in single-qubit gates~\cite{Barends14,Harty14,Ballance16}, two-qubit gates~\cite{Barends14,Ballance16,Rol19a}, and qubit measurements~\cite{Jeffrey14,Harty14,Bultink16,Heinsoo18} at or below the threshold for popular QEC schemes such as surface~\cite{Raussendorf07,Fowler12} and color codes~\cite{Bombin07}. 
They therefore seem well poised for the experimental pursuit of quantum fault tolerance.  
However, a central assumption of textbook QEC, that error processes can be discretized into bit flips ($X$), phase flips ($Z$) or their combination ($Y=iXZ$) only, is difficult to satisfy experimentally.  
This is due to the prevalent use of many-level systems as effective qubits, such as hyperfine levels in ions and weakly anharmonic transmons in superconducting circuits, making leakage from the two-dimensional computational space of effective qubits a threatening error source. 
In quantum dots and trapped ions, leakage events can be as frequent as qubit errors~\cite{Brown18,Andrews19}. However, even when leakage is less frequent than qubit errors as in superconducting circuits~\cite{Barends14,Rol19a}, if ignored, leakage can produce the dominant damage to encoded logical information. 
To address this, theoretical studies propose techniques to reduce the effect of leakage by periodically moving logical information, and removing leakage when qubits are free of logical information~\cite{Aliferis07,Fowler13,Ghosh15,Suchara15}.
Alternatively, more hardware-specific solutions have been proposed for trapped ions~\cite{Brown19} and quantum dots~\cite{Cai19}.
In superconducting circuits, recent experiments have demonstrated single- and  multi-round parity measurements to correct qubit errors with up to 9 physical qubits~\cite{Riste13b,Liu16,Corcoles15,Riste15,Kelly15,Takita16,Takita17,Harper19,Kraglund19}. 
Parallel approaches encoding information in the Hilbert space of single resonators using cat~\cite{Ofek16} and binomial codes~\cite{Hu19} used transmon-based photon-parity checks to approach the break-even point for a quantum memory. 
However, no experiment has demonstrated the ability to detect and mitigate leakage in a QEC context.

\begin{figure*}[ht!]    
\centering      
\includegraphics{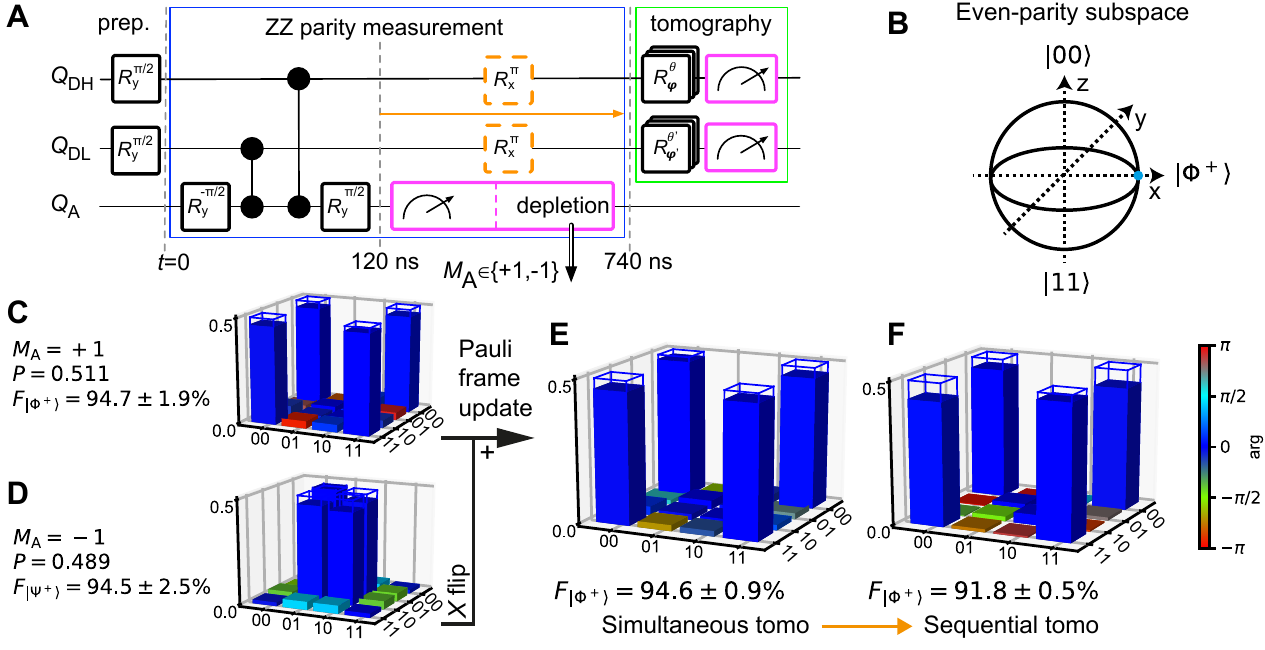}
\caption{\label{fig:ZZ} 
Entanglement genesis by $\SZZ$ parity measurement and Pauli frame update.
(\textbf{A})~Quantum circuit for a parity measurement of the data qubits via coherent operations with ancilla $\QA$ and $\QA$ measurement.  
Tomography reconstructs the data-qubit output density matrix ($\rho$).
Echo pulses (orange) are applied halfway the $\QA$ measurement when performing tomography sequential to the $\QA$ measurement.
(\textbf{B})~Bloch-sphere representation of the even-parity subspace with a marker on $\ket{\Phi^+}$.
(\textbf{C} to \textbf{F})~Plots of $\rho$ with fidelity to the Bell states (indicated by frames) for tomography simultaneous with $\QA$ measurement (C to E) and sequential to $\QA$ measurement (F). 
(C)[(D)]~Conditioning on $\MA=+1[-1]$ ideally generates $\ket{\Phi^+}$ [$\ket{\Psi^+}$] with equal probability $P$. 
(E)[(F)]~PFU applies bit-flip correction ($X$ on $\QDH$) for $\MA=-1$ and reconstructs $\rho$ using all data for simultaneous [sequential] tomography.   
	} 
\end{figure*}

\begin{figure}[ht!]   
\centering     
\includegraphics{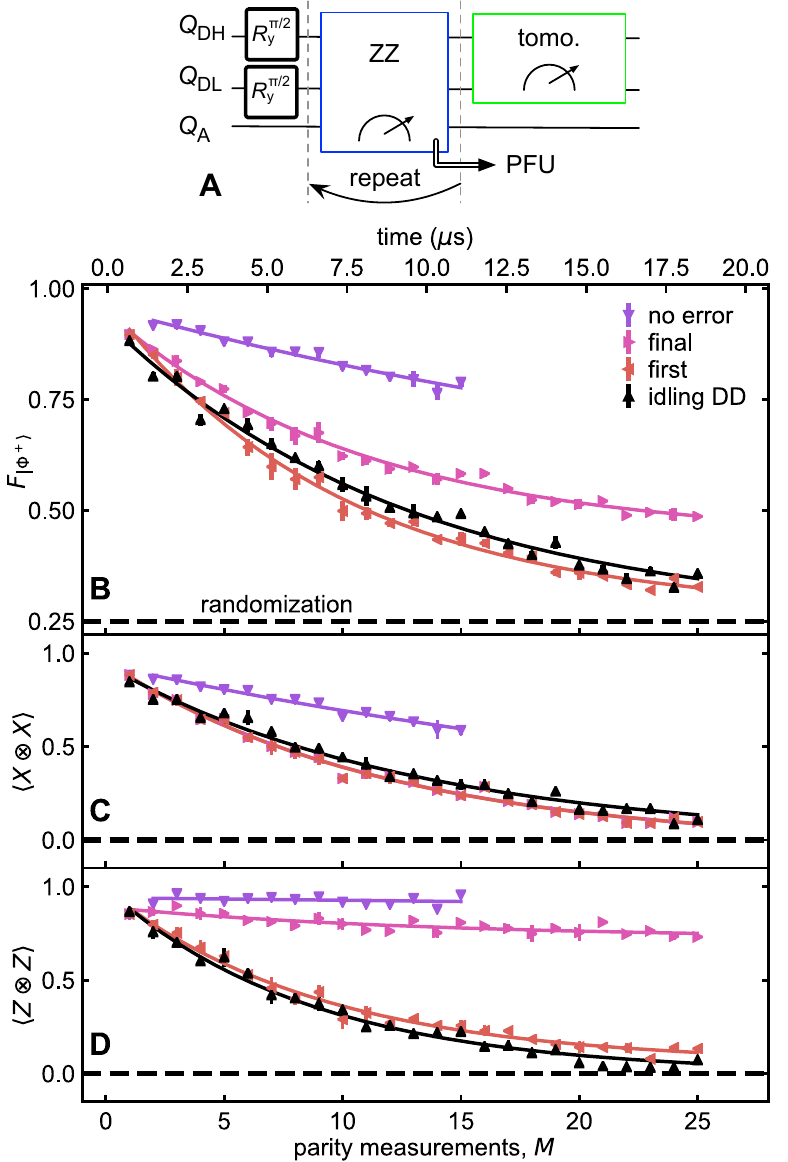}
\caption{\label{fig:ZZs} 
Protecting entanglement from bit flips with repeated $\SZZ$ checks. 
(\textbf{A})~The quantum circuit of \cref{fig:ZZ} A extended with $M$ rounds of repeated $\SZZ$ checks. 
(\textbf{B})~Fidelity to $\ket{\Phi^+}$ as a function of $M$. 
`No error' postselects the runs in which no bit flip is detected. 
`Final' applies PFU based on the last two outcomes (equivalent to minimum-weight perfect matching).
`First' uses the first parity outcome only. 
`Idling DD' are Bell states evolving under dynamical decoupling only (quantum circuit in \cref{fig:idling}). 
(\textbf{C})~Corresponding $\XX$. `final' coincides with `first'. 
(\textbf{D})~Corresponding $\ZZ$. The weak degradation observed for `final' is the hallmark of leakage.
Curves in (B to D) are best fits of a simple exponential decay. 
} \end{figure}

\begin{figure*}[ht!]   
\centering     
\includegraphics{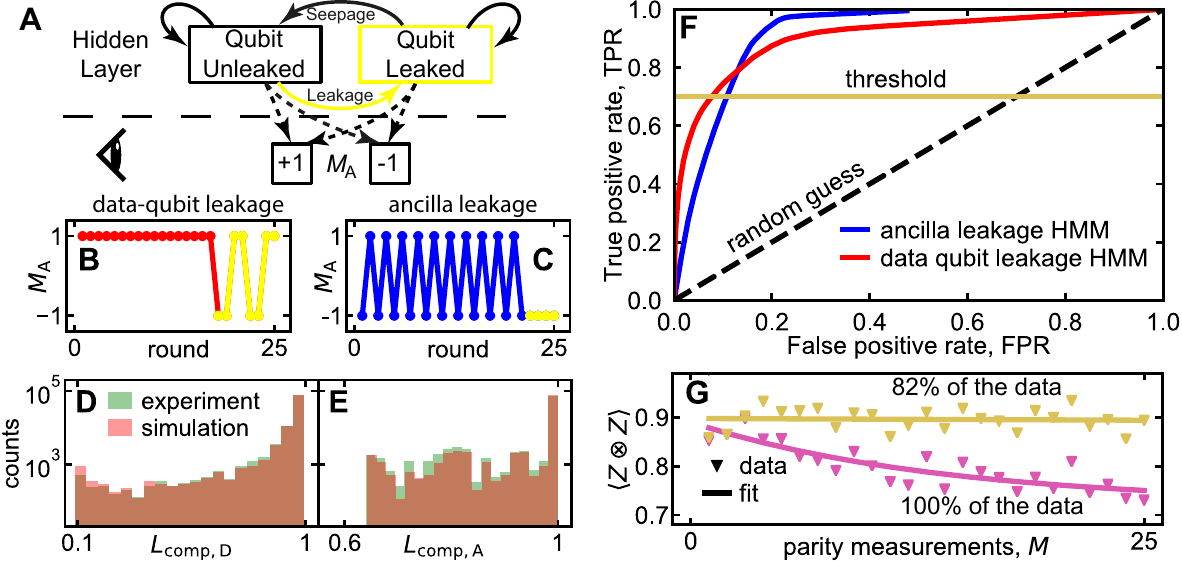}
\caption{\label{fig:leakage}   
Leakage detection and mitigation during repeated $\SZZ$ checks using hidden Markov models (HMMs). 
(\textbf{A})~Simplified HMM.  
In each round, a hidden state (leaked or unleaked)  (top) is updated probabilistically (full arrows), and produces an observable  $\MA$ (bottom) with state-dependent probabilities (dashed arrows). 
After training, the HMM  can be used to assess the likelihood of states given a produced string $\vecs$ of $\MA$.
(\textbf{B})~Example $\vecs$ for a data-qubit leakage event (yellow markers), showing the characteristic pattern of repeated errors.
(\textbf{C})~Example $\vecs$ for $\QA$ leakage signalled by constant $\MA=-1$.
(\textbf{D})~Histograms of $10^5$ $\vecs$ with $M=25$, obtained both experimentally, and simulated by the HMM optimized to detect data-qubit leakage, binned according to the likelihood (\cref{eq:lcompD_def,eq:lcompA_def}) of the data qubits being unleaked (as assessed from the trained HMM). 
HMM training suggests $5.6\%$ total data-qubit leakage at $M=25$ [calculated from \cref{tab:HMM_params} as the steady-state fraction $p_{\mathrm{leak}}/(p_{\mathrm{leak}}+p_{\mathrm{seep}})]$. 
(\textbf{E})~Corresponding histograms using the HMM optimized for $\QA$ leakage. 
This HMM suggests $3.8\%$ total $\QA$ leakage.
(\textbf{F})~Receiver operating characteristics for the trained HMMs.  
(\textbf{G})~$\ZZ$ after $M$ $\SZZ$ checks and correction based on the `final' outcomes, without (same data as in \cref{fig:ZZs}D) and with leakage mitigation by postselection ($\TPR=0.7$).
} \end{figure*}

\begin{figure}[ht!]   
\centering     
\includegraphics{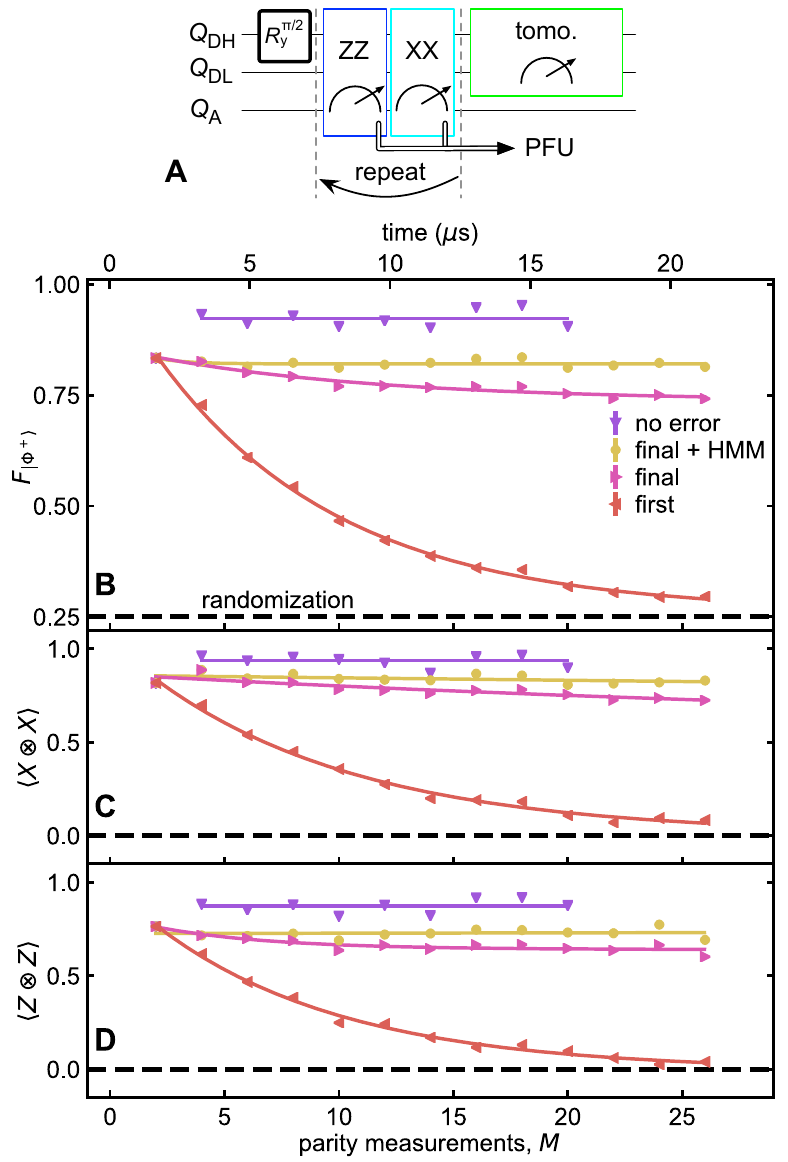}
\caption{\label{fig:ZZXXs} 
Protecting entanglement from general qubit error and leakage.
(\textbf{A})~Simplified quantum circuit with preparation, repeated pairs of $\SZZ$ and $\SXX$ checks, and data-qubit tomography. 
(\textbf{B})~Fidelity to $\ket{\Phi^+}$ as a function of $M$, extracted from the data-qubit tomography. 
`No error' postselects the runs in which no error is detected (postselected fraction in \cref{fig:fractions}).
`Final' applies PFU based on the last three outcomes (equivalent to mimimum-weight perfect matching).
`Final + HMM' includes mitigation of leakage.
`First' uses only the first pair of parity outcomes. 
(\textbf{C} and \textbf{D})~Corresponding $\XX$ and $\ZZ$.
Curves in (B to D) are best fits of a simple exponential decay. 
} \end{figure}

In this report, we experimentally investigate leakage detection and mitigation in a minimal QEC system. 
Specifically, we protect an entangled state of two transmon data qubits ($\QDH$ and $\QDL$) from qubit errors and leakage during up to $26$ rounds of parity measurements via an ancilla transmon ($\QA$).
Performing these parity checks in the $Z$ basis protects the state from $X$ errors, while interleaving checks in the $Z$ and $X$ bases protects it from general qubit errors ($X$, $Y$ and $Z$).  
Leakage manifests itself as a round-dependent degradation of data-qubit correlations ideally stabilized by the parity checks: $\ZZ$ in the first case and $\XX$, $\YY$, and $\ZZ$ in the second.  
We introduce hidden Markov models (HMMs) to efficiently detect data-qubit and ancilla leakage, using only the string of parity outcomes, demonstrating restoration of the relevant correlations. 
Although we use postselection here, the low technical overhead of HMMs makes them ideal for real-time leakage correction in larger QEC codes.

\section{Results}
\subsection{A mimimal QEC setup}
\label{sec:2.1}
Repetitive parity checks can produce and stabilize two-qubit entanglement.  
For example, performing a $Z\texttt{$\otimes$}Z$ parity measurement (henceforth a $\SZZ$ check) on two data qubits prepared in the unentangled state $\ket{\texttt{++}}=(\ket{0}+\ket{1})\otimes (\ket{0}+\ket{1})/2$ will ideally project them to either of the two (entangled) Bell states $\ket{\Phi^+}=(\ket{00}+\ket{11})/\sqrt{2}$ or $\ket{\Psi^+}=(\ket{01}+\ket{10})/\sqrt{2}$, as signaled by the ancilla measurement outcome $\MA$. 
Further $\SZZ$ checks will ideally leave the entangled state unchanged. 
However, qubit errors will alter the state in ways that may or may not be detectable and/or correctable.  
For instance, a bit-flip ($X$) error on either data qubit, which transforms $\ket{\Phi^+}$ into $\ket{\Psi^+}$, will be detected because $X$ anti-commutes with a $\SZZ$ check. 
The corruption can be corrected by applying a bit flip on either data qubit because this cancels the original error ($X^2=I$) or completes the operation $X\texttt{$\otimes$} X$, of which $\ket{\Phi^+}$ and $\ket{\Psi^+}$ are both eigenstates. 
The correction can be applied in real time using feedback~\cite{Riste13b,Liu16,Negnevitsky18,Kraglund19} or kept track of using Pauli frame updating (PFU)~\cite{Knill05,Kelly15}. 
We choose the latter, with PFU strategy "$X$ on $\QDH$". 
Phase-flip errors are not detectable since $Z$ on either data qubit commutes with a $\SZZ$ check. 
Such errors transform $\ket{\Phi^+}$ into $\ket{\Phi^-}=(\ket{00}-\ket{11})/\sqrt{2}$ and $\ket{\Psi^+}$ into $\ket{\Psi^-}=(\ket{01}-\ket{10})/\sqrt{2}$. 
Finally, $Y$ errors produce the same signature as $X$ errors. 
Our PFU strategy above converts them into $Z$ errors. 
Crucially, by interleaving checks of type $\SZZ$ and $\SXX$ (measuring $X\texttt{$\otimes$} X$), arbitrary qubit errors can be detected and corrected. The $\SZZ$ check will signal either $X$ or $Y$ error, and the $\SXX$ check will signal $Z$ or $Y$, providing a unique signature in combination.

\subsection{Generating entanglement by measurement}
\label{sec:2.2}
Our parity check is an indirect quantum measurement involving coherent interactions of the data qubits with $\QA$ and subsequent $\QA$ measurement~\cite{Saira14} (\cref{fig:ZZ}A). 
The coherent step maps the data-qubit parity onto $\QA$ in $120$~ns using single-qubit (SQ) and two-qubit controlled-phase (CZ) gates~\cite{Rol19a}.
Gate characterizations~\cite{Suppmaterial} indicate state-of-the-art gate errors $e_{\mathrm{SQ}}=\{0.08\pm0.02,0.14\pm0.016,0.21\pm0.06\}\%$ and $e_{\mathrm{CZ}}=\{1.4\pm0.6, 0.9\pm0.16\}\%$ with leakage per CZ $L_1=\{0.27\pm0.12, 0.15\pm0.07\}\%$. 
We measure $\QA$ with a $620$-ns pulse including photon depletion~\cite{McClure16,Bultink16}, achieving an assignment error $e_{\mathrm{a}}=1.0\pm0.1\%$. 
We avoid data-qubit dephasing during the $\QA$ measurement by coupling each qubit to a dedicated readout resonator and a dedicated Purcell filter~\cite{Heinsoo18} (\cref{fig:device}). 
The parity check has a cycle time of $740$~ns, corresponding to only $2.5\pm0.2\%$ and $5.0\pm0.3\%$ of the data-qubit echo dephasing times~\cite{Suppmaterial}.

The parity measurement performance can be quantified by correlating its outcome with input and output states. 
We first quantify the ability to distinguish even- ($\ket{00}, \ket{11}$) from odd-parity ($\ket{01}, \ket{10}$) data-qubit input states, finding an average parity assignment error $e_{\mathrm{a, ZZ}}=5.1\pm0.2\%$. 
Second, we assess the ability to project onto the Bell states by performing a $\SZZ$ check on $\ket{\texttt{++}}$ and reconstructing the most-likely physical data-qubit output density matrix $\rho$, conditioning on $\MA=\pm 1$. 
When tomographic measurements are performed simultaneously with the $\QA$ measurement, we find Bell-state fidelities $F_{\ket{\Phi^+}|\MA=+1} = \bra{\Phi^+}\rho_{\MA=+1}\ket{\Phi^+}=94.7\pm1.9\%$ and $F_{\ket{\Psi^+}|\MA=-1}=94.5\pm2.5\%$ (\cref{fig:ZZ}, C and D).
We connect $\ket{\Psi^+}$ to $\ket{\Phi^+}$ by incorporating the PFU into the tomographic analysis,
obtaining $\FBell=94.6\pm0.9\%$ without any postselection (\cref{fig:ZZ}E). 
The nondemolition character of the $\SZZ$ check is then validated by performing tomography only once the $\QA$ measurement completes. 
We include an echo pulse on both data qubits during the $\QA$ measurement to reduce intrinsic decoherence and negate residual coupling between data qubits and $\QA$ (\cref{fig:dephasing}).
The degradation to $\FBell=91.8\pm0.5\%$ is consistent with intrinsic data-qubit decoherence under echo and confirms that measurement-induced errors are minimal.

\subsection{Protecting entanglement from bit flips and the observation of leakage}
\label{sec:2.3}
QEC stipulates repeated parity measurements on entangled states.
We therefore study the evolution of $\FBell=(1+\XX-\YY+\ZZ)/4$ and its constituent correlations as a function of the number $M$ of checks (\cref{fig:ZZs}A). When performing PFU using the first $\SZZ$ outcome only (ignoring subsequent outcomes), we observe that $\FBell$ witnesses entanglement ($>0.5$) during 10 rounds and approaches randomization ($0.25$) by $M=25$ (\cref{fig:ZZs}B).
The constituent correlations also decay with simple exponential forms. A best fit of the form $\ZZ[M]=a\cdot e^{-M/\upsilon_{ZZ}}+b$ gives a decay time $\upsilon_{\mathrm{ZZ}}=9.0\pm0.9$ rounds; similarly, we extract $\upsilon_{\mathrm{XX}}=11.7\pm1.0$ rounds (\cref{fig:ZZs}, C and D). 
By comparison, we observe that Bell states evolving under dynamical decoupling only (no $\SZZ$ checks, see \cref{fig:idling}) decay similarly ($\upsilon_{\mathrm{ZZ}}=8.6\pm0.3$, $\upsilon_{\mathrm{XX}}=12.8\pm0.4$~rounds). 
These similarities indicate that intrinsic data-qubit decoherence is also the dominant error source in this multi-round protocol.

To demonstrate the ability to detect $X$ and $Y$ but not $Z$ errors, we condition the tomography on signaling no errors during $M$ rounds. 
This boosts $\ZZ$ to a constant, while the undetectability of $Z$ errors only allows slowing the decay of $\XX$ to $\upsilon_{\mathrm{XX}}=33.2\pm1.7$~rounds (and of $\YY$ to $\upsilon_{\mathrm{YY}}=31.3\pm1.9$~rounds). 
Naturally, this conditioning comes at the cost of the postselected fraction $\fpost$ reducing with $M$ (\cref{fig:fractions}).

Moving from error detection to correction, we consider the protection of $\Lplus$ by tracking $X$ errors and applying corrections in post-processing.
The correction relies on the final two $\MA$ only, concluding even parity for equal measurement outcomes and odd parity for unequal. 
For this small-scale experiment, this strategy is equivalent to a decoder based on minimum-weight perfect matching (MWPM)~\cite{Fowler12,oBrien17}, justifying its use. 
Because our PFU strategy converts $Y$ errors into $Z$ errors, one expects a faster decay of $\XX$ compared to the no-error conditioning;  indeed, we observe $\upsilon_{\mathrm{XX}}=11.8\pm1.0$~rounds.
Most importantly, correction should lead to a constant $\ZZ$. 
While $\ZZ$ is clearly boosted, a weak decay to a steady state $\ZZ=0.73\pm0.03$ is also evident (\cref{fig:ZZs}D). 
As previously observed in Ref.~\cite{Negnevitsky18}, this degradation is the hallmark of leakage [see also~\cite{Kelly15,Liu16}].
We additionally compare the experimental results to simulations using a model that assumes ideal two-level systems~\cite{oBrien17} (no leakage) based on independently calibrated parameters of ~\cref{tab:device_parameters} (\cref{fig:no_leakage_model} A to D). 
At $M=1$ model and experiment coincide for all correction strategies. 
At larger $M$ `first' and `final' correction strategies deviate significantly, consistent with a gradual build-up of leakage, which we now turn our focus to.

\subsection{Leakage detection using hidden Markov models}
\label{sec:2.4}
Both ancilla and data-qubit leakage in our experiment can be inferred from a string $\vecs=(\MA[m=0],\ldots,\MA[m=M])$ of measurement outcomes.
Leakage of $\QA$ to the second excited transmon state $\ket{2}$ produces $\MA=-1$ because measurement does not discern it from $\ket{1}$.
This leads to the pattern $\vecs=(\texttt{$\ldots$}-1,-1\texttt{$\ldots$})$ until $\QA$ seeps back to $\ket{1}$ (coherently or by relaxation), as it is unaffected by subsequent $\pi/2$ rotations (\cref{fig:leakage}C).
Leakage of a data qubit (\cref{fig:leakage}B) leads to apparent repeated errors [signaled by $\vecs=(\texttt{$\ldots$}+1,+1,-1,-1\texttt{$\ldots$})$], as the echo pulses only act on the unleaked qubit.
This is equivalent to a pattern of repeated error signals in the data-qubit syndrome $\sD[m]:=\MA[m]\cdot\MA[m-2]$ --- $\sD=(\texttt{$\ldots$},-1,-1,-1,\texttt{$\ldots$})$.
(We call $\sD[m]=-1$ an error signal as in the absence of noise $\sD[m]=+1$, while the measurements $\MA[m]$ will still depend on the $\SZZ$ parity.)

Neither of the above patterns is entirely unique to leakage; each may also be produced by some combination of qubit errors.
Therefore, we cannot unambiguously diagnose an individual experimental run of corruption by leakage.
However, given a set of ancilla measurements $\MA[0],\ldots,\MA[m]$, the likelihood $\Lcomp(\vecs)$ that qubit $Q$ is in the computational subspace during the final parity checks is well-defined.
In this work, we infer $\Lcomp(\vecs)$ by using a hidden Markov model (HMM)~\cite{Baum66}, which treats the system as leaking out of and seeping back to the computational subspace in a stochastic fashion between each measurement round (a leakage HMM in its simplest form is shown in \cref{fig:leakage}A, and further described in \cref{sec:MM.HMM.general,sec:MM.HMM.forQEC,sec:MM.HMM.twostate}).
This may be extended to scalable leakage detection (for the purposes of leakage mitigation) in a larger QEC code, by using a separate HMM for each data qubit and ancilla.
To improve the validity of the HMMs, we extend their internal states to allow the modeling of additional noise processes in the experiments (detailed in \cref{sec:MM.HMM.channels,sec:MM.HMM.models}).

Before assessing the ability of our HMMs to improve fidelity in a leakage mitigation scheme, we first validate and benchmark them internally.
A common method to validate the HMM's ability to model the experiment is to compare statistics of the experimentally-generated data to a simulated data set generated by the model itself. 
As we are concerned only with the ability of the HMM to discriminate leakage, $\Lcomp(\vecs)$ provides a natural metric for comparison.
In \cref{fig:leakage}, D and E, we overlay histograms of $10^5$ experimental and simulated experiments, binned according to $\Lcomp(\vecs)$, and observe excellent agreement.
To further validate our model, we calculate the Akaike information criterion~\cite{Akaike74}:
\begin{equation}
\label{eq:akaike}
A(H)=2n_{\mathrm{p},H}-2\log\left[\max_{p_i}L(\{\vec{o}\}|H\{p_i\})\right],
\end{equation}
where $L(\vec{o}|M)$ is the likelihood of making the set of observations $\{\vec{o}\}$ given model $H$ (maximized over all parameters $p_i$ in the model, as listed in \cref{tab:HMM_params}.), and $n_{\mathrm{p},M}$ is the number of parameters $p_i$.
The number $A(H)$ is rather meaningless by itself; we require a comparison model $H^{(\mathrm{comp})}$ for reference.
Our model is preferred over the comparison model whenever $A(H)>A(H^{(\mathrm{comp})})$.
For comparison, we take the target HMM $H$, remove all parameters describing leakage, and re-optimize.
We find the difference $A(H)-A(H^{(\mathrm{comp})})=1.1\times 10^5$ for the data-qubit HMM, and $2.1\times 10^4$ for the ancilla HMM, giving significant preference for the inclusion of leakage in both cases.
[The added internal states beyond the simple two-state HMMs clearly improves the overlap in histograms, \cref{fig:simple_model_performance}, A and B. The added complexity is further justified by the Akaike information criterion~\cite{Suppmaterial}].

The above validation suggests that we may assume that the ratio of actual leakage events at a given $\Lcomp$ is well approximated by $\Lcomp$ itself (which is true for the simulated data).
Under this assumption, we expose the HMMs discrimination ability by plotting its receiver operating characteristic~\cite{Green66} (ROC).
The ROC (\cref{fig:leakage}F) is a parametric plot (sweeping a threshold $\Lcompth$) of the true positive rate $\TPR$ (the fraction of leaked runs correctly identified) versus the false positive rate $\FPR$ (the fraction of unleaked runs wrongly identified). 
Random rejection follows the line $y=x$; the better the detection the greater upward shift. 
Both ROCs indicate that most of the leakage ($\TPR=0.7$) can be efficiently removed  with $\FPR \sim 0.1$. 
Individual mappings of $\TPR$ and $\FPR$ as a function of $\Lcompth$ can be found in \cref{fig:L_tradeoff}, A and B.
Further rejection is more costly, which we attribute to these leakage events being shorter-lived. 
This is because the shorter a leakage event, the more likely its signature is due to (a combination of) qubit errors.
Fortunately, shorter leakage events are also less damaging. 
For instance, a leaked data qubit that seeps back within the same round may be indistinguishable from a relaxation event, but also has the same effect on encoded logical information~\cite{Fowler13}.

We now verify and externally benchmark our HMMs by their ability to improve $\ZZ$ by rejecting data with a high probability of leakage.
To do this, we set a threshold $\Lcompth$, and reject experimental runs whenever $\Lcomp(\vecs)<\Lcompth$.
For both HMMs we choose $\Lcompth$ to achieve $\TPR=0.7$.
With this choice, we observe a restoration of $\ZZ$ to its first-round value across the entire curve (\cref{fig:leakage}G), mildly reducing $\fpost$ to $0.82$ (averaged over $M$).
This restoration from leakage is confirmed by the `final + HMM' data matching the no-leakage model results in \cref{fig:no_leakage_model}, A to D.
As low $\Lcomp(\vecs)$ is also weakly correlated with qubit errors, the gain in $\ZZ$ is partly due to false positives. 
Of the $\sim0.13$ increase at $M=25$, we attribute $0.07$ to actual leakage (estimated from the ROCs). 
By comparison, the simple two-state HMM, leads to a lower improvement, whilst rejecting a larger part of the data (\cref{fig:simple_model_performance}G), ultimately justifying the increased HMM complexity in this particular experiment.

\subsection{Protecting entanglement from general qubit errors and mitigation of leakage}
\label{sec:2.5}
We finally demonstrate leakage mitigation in the more interesting scenario where $\Lplus$ is protected from general qubit error by interleaving $\SZZ$ and $\SXX$ checks~\cite{Negnevitsky18,Kraglund19}. $\SZZ$ may be converted to $\SXX$ by adding $\pi/2$ $y$ rotations on the data qubits simultaneous with those on $\QA$.
This requires that we change the definition of the syndrome to $\sD[m]=\MA[m]\cdot\MA[m-1]\cdot\MA[m-2]\cdot\MA[m-3]$, as we need to `undo' the interleaving of the $\SZZ$ and $\SXX$ checks to detect errors.
For an input state $\ket{\texttt{+}0}=(\ket{0}+\ket{1})/\sqrt{2} \otimes \ket{0}$, a first pair of checks ideally projects the data qubits to one of the four Bell states with equal probability. 
Expanding the PFU to $X$ and/or $Z$ on $\QDH$ we find $\FBell=83.8\pm0.8\%$ (\cref{fig:ZZXX}). 
For subsequent rounds, the `final' strategy now relies on the final three $\MA$. 
We observe a decay towards a steady state $\FBell=73.7\pm0.9\%$ (\cref{fig:ZZXXs}), consistent with previously observed leakage.
We battle this decay by adapting the HMMs (detailed in \cref{sec:MM.HMM.channels,sec:MM.HMM.models}).
We find an improved ROC for $\QA$ leakage (\cref{fig:ROCZZXX}).
For data-qubit leakage however, the ROC is degraded.
This is to be expected --- when one data qubit is leaked in this experiment, the ancilla effectively performs interleaved $Z$ and $X$ measurements on the unleaked qubit.
This leads to a signal of random noise $P(\sD[m]=-1)=0.5$, which is less distinguishable from unleaked experiments $P(\sD[m]=-1)\sim 0$ than the signal of a leaked data-qubit during the $\ZZ$-only experiment $P(\sD[m]=-1)\sim 1$.
Most importantly, thresholding to $\TPR=0.7$ restores $\XX$ and $\ZZ$, leading to an almost constant $\FBell=82.8\pm0.2\%$ with $\fpost=0.81$ (averaged over $M$), as expected from the no-leakage model results in \cref{fig:no_leakage_model}, E to H. 
In this experiment, the simple two-state HMMs performs almost identically compared to the complex HMM, achieving Bell-state fidelities within $2\%$ whilst retaining the same amount of data (\cref{fig:simple_model_performance}N).

\section{Discussion}
This HMM demonstration provides exciting prospects for leakage detection and correction. 
In larger systems, independent HMMs can be dedicated to each qubit because leakage produces local error signals~\cite{Ghosh15}. 
An HMM for an ancilla only needs its measurement outcomes while a data-qubit HMM only needs the outcomes of the nearest-neighbor ancillas [details in~\cite{Suppmaterial}]. 
Therefore, the computational power grows linearly with the number of qubits, making the HMMs a small overhead when running parallel to MWPM.
HMM outputs could be used as inputs to MWPM, allowing MWPM to dynamically adjust its weights. 
The outputs could also be used to trigger leakage reduction units~\cite{Aliferis07,Fowler13,Ghosh15,Suchara15} or qubit resets~\cite{Magnard18}.

In summary, we have performed the first experimental investigation of leakage detection during repetitive parity checking, successfully protecting an entangled state from qubit errors and leakage in a circuit quantum electrodynamics processor.
Future work will extend this protection to logical qubits, e.g., the 17-qubit surface code~\cite{Tomita14,oBrien17}.
The low technical overhead and scalability of HMMs is attractive for performing leakage detection and correction in real time using the same parity outcomes as traditionally used to correct qubit errors only.

\section{Materials and methods}
\subsection{Device}
Our quantum processor (\cref{fig:device}) follows a three-qubit-frequency extensible layout with nearest-neighbor interactions that is designed for the surface code~\cite{Versluis17}. 
Our chip contains low- and high-frequency data qubits ($\QDL$ and $\QDH$), and an intermediate-frequency ancilla ($\QA$). 
Single-qubit gates around axes in the equatorial plane of the Bloch sphere are performed via a dedicated microwave drive line for each qubit.
Two-qubit interactions between nearest neighbors are mediated by a dedicated bus resonator (extensible to four per qubit) and controlled by individual tuning of qubit transition frequencies via dedicated flux-bias lines~\cite{DiCarlo09}.
For measurement, each qubit is dispersively coupled to a dedicated readout resonator (RR) which is itself connected to a common feedline via a dedicated Purcell resonator (PR). 
The RR-PR pairs allow frequency-multiplexed readout of selected qubits with negligible backaction on untargeted qubits~\cite{Heinsoo18}.

\subsection{Hidden Markov models}\label{sec:MM.HMM.general}
HMMs provide an efficient tool for indirect inference of the state of a system given a set of output data~\cite{Baum66}. A hidden Markov model describes a time-dependent system as evolving between a set of $\numstates$ hidden states $\{h\}$ and returning one of $\numoutputs$ outputs $\{o\}$ at each timestep $m$.
The evolution is stochastic: the system state $H[m]$ of the system at timestep $m$ depends probabilistically on the state $H[m-1]$ at the previous timestep, with probabilities determined by a $\numstates\times\numstates$ transition matrix $A$
\begin{equation}
A_{h,h}=P(H[m]=h\;|\;H[m-1]=h').
\end{equation}
The user cannot directly observe the system state, and must infer it from the outputs $O[m]\in\{o\}$ at each timestep $m$.
This output is also stochastic: $O[m]$ depends on $H[m]$ as determined by a $\numoutputs\times\numstates$ output matrix $B$
\begin{equation}
B_{o,h}=P(O[m]=o\; |\; H[m]=h).
\end{equation}

If the $A$ and $B$ matrices are known, along with the expected distribution $\vec{\pi}^{(\mathrm{prior})}[0]$ of the system state over the $\numstates$ possibilities,
\begin{equation}
\pi^{(\mathrm{prior})}_h[1]=P(H[1]=h),
\end{equation}
one may simulate the experiment by generating data according to the above rules.
Moreover, given a vector $\vec{o}$ of observations, we may calculate the distribution $\vec{\pi}[m]$ over the possible states at a later time $m$,
\begin{equation}
\pi^{(\mathrm{post})}_h[m]=P(H[m]=h|O[1]=o_1,\ldots,O[m]=o_m),
\end{equation}
by interleaving rounds of Markovian evolution,
\setlength{\arraycolsep}{0pt}
\begin{eqnarray}
    &\pi^{(\mathrm{prior})}_h[m]\\
    &:=P(H[m]=n|O[1]=o_1,\ldots,O[m-1]=o_{m-1})\\
    &=\sum_{h'}A_{h,h'}\pi^{(\mathrm{post})}_{h'}[m-1],
\end{eqnarray}
\setlength{\arraycolsep}{10pt}
and Bayesian update,
\begin{equation}
    \pi^{(\mathrm{post})}_h[m]=\frac{B_{o_m,h}\;\pi^{(\mathrm{prior})}_h[m]}{\sum_{h'}B_{o_m,h'}\pi^{(\mathrm{prior})}_{h'}[m]}.
\end{equation}

\subsection{Hidden Markov models for QEC experiments}\label{sec:MM.HMM.forQEC}
To maximize the discrimination ability of HMMs in the various settings studied in this work, we choose different quantities to use for our output vectors $\vec{o}$.
In all experiments in this work, the signature of a leaked ancilla is repeated $\MA[m]=-1$, and so we choose $\vec{o}=\vec{\MA}$.
By contrast, the signature of leaked data qubits in both experiments may be seen as an increased error rate in their corresponding syndromes $\vec{\sD}$, and we choose $\vec{o}=\vec{\sD}$ for the corresponding HMMs.

One may predict the computational likelihood for data-qubit (D) leakage at timestep $M$ in the $\SZZ$-check experiment given $\vec{\pi}(M)$.
In particular, once we have declared which states $h$ correspond to leakage, we may write
\begin{equation}
    L_{\mathrm{comp, D}}=\sum_{h\;\mathrm{unleaked}}\pi^{(\mathrm{post})}_h[M].\label{eq:lcompD_def}
\end{equation}
However, in the repeated $\SZZ$-check experiment, the ancilla (A) needs to be within the computational subspace for two rounds to perform a correct parity measurement.
Therefore, the computational likelihood is slightly more complicated to calculate,
\begin{equation}
    L_{\mathrm{comp, A}}[M]=\frac{\sum_{h, h'\;\mathrm{unleaked}}B_{o_m,h}\;A_{h,h'}\pi^{(\mathrm{post})}_{h'}[M-1]}{\sum_{h,h'}B_{o_m,h}A_{h,h'}\pi^{(\mathrm{post})}_{h'}[M-1]}.\label{eq:lcompA_def}
\end{equation}
In the interleaved $\SZZ$- and $\SXX$-check experiment, the situation is more complicated as we require data from the final two parity checks to fully characterize the quantum state.
This implies that we need unleaked data qubits for the last two rounds and unleaked ancillas for the last three.
The likelihood of the latter may be calculated by similar means to the above.

\subsection{Simplest models for leakage discrimination}\label{sec:MM.HMM.twostate}
One need not capture the full dynamics of the quantum system in a HMM to infer whether a qubit is leaked.
This is of critical importance if we wish to extend this method for the purposes of leakage mitigation in a large QEC code [as we discuss in~\cite{Suppmaterial}].
The simplest possible HMM (\cref{fig:leakage}A) has two hidden states: $H[m]=1$ if the qubit(s) in question are within the computational subspace, and $H[m]=2$ if $\QA$ (or either data qubit) is leaked.
(The labels $1$ and $2$ are arbitrary here, and explicitly have no correlation with the qubit states $|1\rangle$ and $|2\rangle$.)
Then, the $2\times 2$ transition matrix simply captures the leakage and seepage rates of the system in question:
\begin{eqnarray}
A=\left(\begin{array}{cc}1&0\\0&1\end{array}\right)+\pleak\left(\begin{array}{cc}-1&0\\1&0\end{array}\right)\\
+\pseep\left(\begin{array}{cc}0&1\\0&-1\end{array}\right).\label{eq:transition_simplemodel}
\end{eqnarray}
The $2\times 2$ output matrices then capture the different probabilities of seeing output $O[m]=0$ or $O[m]=1$ when the qubit(s) are leaked or unleaked:
\begin{eqnarray}
B=\left(\begin{array}{cc}1&0\\0&1\end{array}\right)+p_{0,1}\left(\begin{array}{cc}-1&0\\1&0\end{array}\right)\\
+p_{1,0}\left(\begin{array}{cc}0&1\\0&-1\end{array}\right).\label{eq:output_simplemodel}
\end{eqnarray}
When studying data-qubit leakage, $p_{0,1}$ simply captures the rate of errors within the computational subspace.
Then, in the repeared $\SZZ$-check experiment, $p_{1,0}$ captures events such as ancilla or measurement errors that cancel the error signal of a leakage event.
However, in the interleaved $\SZZ$ and $\SXX$ experiment, a leaked qubit causes the syndrome to be random, so we expect $p_{1,0}\sim 0.5$.
When studying ancilla leakage, $p_{1,0}$ is simply the probability of $\ket{2}$ state being read out as $\ket{0}$, and is also expected to be close to $0$. 
However, $p_{0,1}\sim 0.5$, as we do not reset $\QA$ or the logical state between rounds of measurement, and thus any measurement in isolation is roughly equally-likely to be $0$ or $1$.
In all situations, we assume that the system begins in the computational subspace --- $\pi_n(0)=\delta_{n,0}$.
With this fixed, we may choose the parameters $\pleak$, $\pseep$, $p_{0,1}$ and $p_{1,0}$ to maximize the likelihood $L(\{\vec{o}\})$ of observing the recorded experimental data $\{o\}$.
(Note that $L(\{\vec{o}\})$ is not the computational likelihood $\Lcomp$.)

\subsection{Modeling additional noise}\label{sec:MM.HMM.channels}
The simple model described above does not completely capture all of the details of the stabilizer measurements $\vecs$.
For example, the data-qubit HMM will overestimate the leakage likelihood when an ancilla error occurs, as this gives a signal with a time correlation that is unaccounted for.
As the signature of a leakage event in a fully fault-tolerant code will be large~\cite{Suppmaterial}, we expect these details to not significantly hinder the simple HMM in a large-scale QEC simulation.
However, this lack of accuracy makes evaluating HMM performance somewhat difficult, as internal metrics may not be so trustworthy.
We also risk overestimating the HMM performance in our experiment, as our only external metrics for success (e.g.,~fidelity) do just as poorly when errors occur near the end of the experiment as they do when leakage occurs.
Therefore, we extend the set of hidden states in the HMMs to account for ancilla and measurement errors, and to allow the ancilla HMM to keep track of the stabilizer state.
To attach physical relevance to the states in our Markovian model, and to limit ourselves to the noise processes that we expect to be present in the system, we generalize \cref{eq:transition_simplemodel,eq:output_simplemodel} to a linearly-parametrized model,
\begin{equation}
    A=A_0+\sum_{\mathrm{err}}p_{\mathrm{err}}D^{(\mathrm{A})}_{\mathrm{err}},\hspace{0.5cm} B=B_0+\sum_{\mathrm{err}}p_{\mathrm{err}}D^{(\mathrm{B})}_{\mathrm{err}}.
\end{equation}
Here, we choose the matrices $D^{(\mathrm{A})}_i$ and $D^{(\mathrm{B})}_i$ such that the error rates $p^{(\mathrm{A})}_i,p^{(\mathrm{B})}_i$ correspond to known physical processes.
(We add the superscripts $(A)$ and $(B)$ here to the $D$ matrices to emphasize that each error channel only appears in one of the two above equations.)

The error generators $D^{(\mathrm{A})}$, $D^{(\mathrm{B})}$ may be identified as derivatives of $A$ with respect to these error rates:
\begin{equation}
    D^{(\mathrm{A})}_i=\frac{\partial A}{\partial p^{(\mathrm{A})}_i},\hspace{2cm}D^{(\mathrm{B})}_i=\frac{\partial B}{\partial p^{(\mathrm{B})}_i}.
\end{equation}
This may be extended to calculate derivatives of the likelihood $L(\{\vec{o}\})$ (or more practically, the log-likelihood) with respect to the various parameters $p_i$.
This allows us to obtain the maximum likelihood model within our parametrization via gradient descent methods (in particular the Newton-CG method), instead of resorting to more complicated optimization algorithms such as the Baum-Welch algorithm~\cite{Baum66}.
All models were averaged over between $10$ and $20$ optimizations using the Newton-CG method in scipy~\cite{Scipy01}, calculating likelihoods, gradients and Hessians over $10,000$-$20,000$ experiments per iteration, and rejecting any failed optimizations.
As the signal of ancilla leakage is identical to the signal for even $\SZZ$ and $\SXX$ parities with ancilla in $\ket{1}$ and no errors, we find that the optimization is unable to accurately estimate the ancilla leakage rate, and so we fix this in accordance with independent calibration to $0.0040$/round using averaged homodyne detection of $\ket{2}$ (making use of a slightly different homodyne voltage for $\ket{1}$ and $\ket{2}$).

\subsection{Hidden Markov models used in \cref{fig:ZZs,fig:ZZXXs}}
\label{sec:MM.HMM.models}

Different Markov models (with independently optimized parameters) were used to optimize ancilla and data-qubit leakage estimation for both the $\SZZ$ experiment and the experiment interleaving $\SZZ$ and $\SXX$ checks.
This lead to a total of four HMMs, which we label $\HZZ$-D, $\HZZ$-A, $\HZZXX$-D and $\HZZXX$-A.
A complete list of parameter values used in each HMM is given in \cref{tab:HMM_params}. 
We now describe the features captured by each HMM.
As we show in~\cite{Suppmaterial}, these additional features are not needed to increase the error mitigation performance of the HMMs, but rather to ensure their closeness to the experiment and increase trust in their internal metrics.

To go beyond the simple HMM in the $\SZZ$-check experiment when modeling data-qubit leakage ($\HZZ$-D), we need to include additional states to account for the correlated signals of ancilla and readout error.
If we assume data-qubit errors (that remain within the logical subspace) are uncorrelated in time, they are already well-captured in the simple model.
This is because any single error on a data qubit may be decomposed into a combination of $Z$ errors (which commute with the measurement and thus are not detected) and $X$ errors (which anti-commute with the measurement and thus produce a single error signal $\sD[m]=1$), and is thus captured by the $p_{0,1}$ parameter.
When one of the data qubits is leaked, uncorrelated $X$ errors on the other data qubit cancel the constant $\sD[m]=-1$ signal for a single round, and are thus captured by the $p_{1,0}$ parameter.
However, errors on the ancilla, and readout errors, give error signals that are correlated in time (separated by $1$ or $2$ timesteps, respectively).
This may be accounted for by including extra `ancilla error states'.
These may be most easily labeled by making the $h$ labels a tuple $h=(h_0,h_1)$, where $h_0$ keeps track of whether or not the qubit is leaked, and $h_1=1,2,3$ keeps track of whether or not a correlated error has occurred.
In particular, we encode the future syndrome for 2 cycles in the absence of error on $h_1$, allowing us to account for any correlations up to 2 rounds in the future.
This extends the model to a total of $4\times 2 = 8$ states.
The transition and output matrices in the absence of error for the unleaked $h_0=0$ states may then be written in a compact form (noting that leakage errors cancel out with correlated ancilla and readout errors to give $\sD[m]=+1$), 
\begin{equation}
[A_0]_{(h_0,h_1//2),(h_0,h_1)}=1,\hspace{0.5cm} [B_0]_{-1^{h_0+h_1},(h_0,h_1)}=1,\label{eq:A0B0}
\end{equation}
where the double slash $//$ refers to integer division.

Let us briefly demonstrate how the above works for ancilla error in the system.
Suppose the system was in the state $h=(0,3)$ at time $m$.
It would output $\MA[m]=-1$, and then evolve to $h=(0,3//2)=(0,1)$ at time $m+1$ (in the absence of additional error).
Then, it would output a second error signal [$\MA[m+1]=-1$] and finally decay back to the $h=(0,1//2)=(0,0)$ state.
This gives the HMM the ability to model ancilla error as an evolution from $h=(0,0)$ to $h(0,3)$.
Formally, we assign the matrix $D^{(A)}_{\mathrm{ancilla}}$ to this error process, and following this argument we have
\begin{equation}
[D^{(A)}_{\mathrm{ancilla}}\,]_{(0,0),(0,0)}=-1,\;\; [D^{(A)}_{\mathrm{ancilla}}\,]_{(0,3),(0,0)}=1.
\end{equation}
The corresponding error rate $p_{\mathrm{ancilla}}$ is then an additional free parameter to be optimized to maximize the likelihood.
To finish the characterization of this error channel, we need to consider the effect of ancilla error in states other than $h=(0,0)$.
Two ancilla errors in the same timestep cancel, but two ancilla errors in subsequent timesteps will cause the signature $\sD=\ldots,-1,+1,-1,\ldots$.
This may be captured by an evolution from $h=(0,2)$ to $h=(0,3)$ [instead of $h=(0,1)$], which implies we should set 
\begin{equation}
[D^{(A)}_{\mathrm{ancilla}}\,]_{(0,1),(0,3)}=-1,\;\; [D^{(A)}_{\mathrm{ancilla}}\,]_{(0,2),(0,3)}=1.
\end{equation}
(Note that $A_0$ already captures a decay from $h=(0,2)\rightarrow(0,1)\rightarrow(0,0)$, which will give the desired signal.)
We note that this also matches the signature of readout error, which can then be captured by a separate error channel $D^{(A)}_{\mathrm{readout}}$ which increases this correlation
\begin{equation}
[D^{(A)}_{\mathrm{readout}}\,]_{(0,1),(0,2)}=-1,\;\; [D^{(A)}_{\mathrm{readout}}\,]_{(0,3),(0,2)}=1.
\end{equation}
One can then check that ancilla errors in $h=(0,3)$ should cause the system to remain in $h=(0,3)$, and that ancilla or readout errors in $h=(0,1)$ should evolve the system to $h=(0,2)$. We note that this model cannot account for the $\sD=\ldots-1,+1,+1,+1,-1$ signature of readout error at time $m$ and $m+2$, but adjusting the model to include this has negligible effect.

Ancilla error in the~$\SZZ$-check experiment when the data qubits are leaked has the same correlated behavior as when they are not, but may occur at a different rate.
This requires that we define a new matrix $D^{(A)}_{\mathrm{ancilla, leaked}}$ by
\begin{equation}
[D^{(A)}_{\mathrm{ancilla, leaked}}]_{(1,j),(1,k)}=[D^{(A)}_{\mathrm{ancilla}}]_{(0,j),(0,k)},
\end{equation}
with a separate error rate $p_{\mathrm{ancilla, leaked}}$.
As we do not expect the readout of the ancilla to be significantly affected by whether the data qubit is leaked, we do not add an extra parameter to account for this behavior, and instead simply set
\begin{equation}
[D^{(A)}_{\mathrm{readout}}]_{(1,j),(1,k)}=[D^{(A)}_{\mathrm{readout}}]_{(0,j),(0,k)}.
\end{equation}
We also assume that leakage $p_{\mathrm{leak}}$ and seepage $p_{\mathrm{seep}}$ rates are independent of these correlated errors (i.e., $[D^{(A)}_{\mathrm{leak}}]_{(0,j),(0,k)},[D^{(A)}_{\mathrm{seep}}]_{(1,j),(1,k)}]\in\{0,-1\}$).
We then assume that the first measurement made following a leakage/seepage event is just as likely to have an additional error (corresponding to an evolution to $(h_0,1)$) or not (corresponding to an evolution to $(h_0,0)$).
We finally account for data-qubit error in the output matrices in the same way as in the simple model, but with different error rates $p_{\mathrm{data, leaked}}$ for the leaked states $(1,h_1)$ and $p_{\mathrm{data}}$ for the unleaked states $(0,h_1)$.

There are a few key differences between the interleaved $\SZZ$---$\SXX$ and $\SZZ$ experiments that need to be captured in the data-qubit HMM $\HZZXX-D$.
Firstly, as the syndrome is now given by $\sD[m]=\MA[m]\cdot\MA[m-1]\cdot\MA[m-2]\cdot\MA[m-3]$, ancilla and classical readout error can then generate a signal stretching up to $4$ steps in time.
This implies that we require $2^4$ possibilities for $h_1$ to keep track of all correlations.
However, as a leaked data qubit makes ancilla output random in principle, we no longer need to keep track of the ancilla output upon leakage.
This implies that we can accurately model the experiment with $16+1=17$ states, which we can label by $h\in\{2,(1,h_1)\}$.
The $A_0$ and $B_0$ matrices in the unleaked states $(1,h_1)$ follow \cref{eq:A0B0}, and we fix $[A_0]_{2,2}=1$ (as in the absence of $p_{\mathrm{seep}}$ a leaked state stays leaked).
However, we allow for some bias in the leaked state error rate - $B_{-1,2}=p_{\mathrm{data, leaked}}$ is not fixed to $0.5$.
(For example, this accounts for a measurement bias towards a single state, which will reduce the error rate below $0.5$.)
The non-zero elements in the matrices $D^{(A)}_{\mathrm{ancilla}}$ and $D^{(A)}_{\mathrm{readout}}$ may be written:
\begin{eqnarray}
{}[D^{(A)}_{\mathrm{ancilla}}]_{(1,h_1//2),(1,h_1)}=&-1,\\
{}[D^{(A)}_{\mathrm{ancilla}}]_{(1,h_1//2\oplus 5)}=&1,\\
{}[D^{(A)}_{\mathrm{readout}}]_{(1,h_1//2),(1,h_1)}=&-1,\\
{}[D^{(A)}_{\mathrm{readout}}]_{(1,h_1//2\oplus 15)}=&1.
\end{eqnarray}
Here, $a\oplus b$ refers to addition of each binary digit of $a$ and $b$ modulo $2$.
We may use this formalism to additionally keep track of $Y$ data-qubit errors, which show up as correlated errors on subsequent $\SXX$ and $\SZZ$ stabilizer checks, by introducing a new error channel
\begin{equation}
[D^{(A)}_{\mathrm{data,Y}}]_{(1,h_1//2),(1,h_1)}=-1,\hspace{0.5cm}[D^{(A)}_{\mathrm{data,Y}}]_{(1,h_1//2\oplus 3)}=1,
\end{equation}
with a corresponding error rate $p_{\mathrm{data,Y}}$.
As before, we assume that leakage occurs at a rate $p_{\mathrm{leak}}$ independently of $h_1$, and that seepage takes the system either to the state with either no error signal $h=(1,0)$ or one error signal $h=(0,1)$ with a rate $p_{\mathrm{seep}}$.

As the output used for the $\HZZ-A$ HMM is the pure measurement outcomes $\MA$, the dominant signal that must be accounted for is that of the stabilizer $\SZZ$ itself.
This either causes a constant signal $\MA[m]=\MA[m-1]$ or a constant flipping signal $\MA[m]=-\MA[m-1]$.
This cannot be accounted for in the simple HMM, as it cannot contain any history in a single unleaked state.
To deal with this, we extend the set of states in the $\HZZ-A$ HMM to include both an estimate of the ancilla state $a\in\{0,1,2\}$ at the point of measurement, and the stabilizer state $s\in\{0,1\}$, and label the states by the tuple $(a,s)$.
The ancilla state then immediately defines the device output in the absence of any error:
\begin{equation}
[B_0]_{1,(0,s)}=[B_0]_{-1,(1,s)}=[B_0]_{-1,(2,s)}=1,
\end{equation}
while the stabilizer state defines the transitions in the absence of any error or leakage:
\begin{equation}
[A_0]_{(a+s\;\mathrm{mod}\;2,s)(a,s)}=1\; \mathrm{if}\;a<2,\hspace{1cm}[A_0]_{(2,s),(2,s)}=1.
\end{equation}
The only thing that affects the output matrices is readout error:
\begin{eqnarray}
&[D_{\mathrm{readout}}^{(B)}]_{1,(0,s)}\\
&=[D_{\mathrm{readout}}^{(B)}]_{-1,(1,s)}=[D_{\mathrm{readout}}^{(B)}]_{-1,(2,s)}=-1,\\
&[D_{\mathrm{readout}}^{(B)}]_{-1,(0,s)}\\
&=[D_{\mathrm{readout}}^{(B)}]_{1,(1,s)}=[D_{\mathrm{readout}}^{(B)}]_{1,(2,s)}=1.
\end{eqnarray}
Data-qubit errors flip the stabilizer with probability $p_{\mathrm{data}}$:
\begin{eqnarray}
[D_{\mathrm{data}}^{(A)}]_{(a,s),(a',s)}=&-[A_0]_{(a,s),(a',s)},\\
{}[D_{\mathrm{data}}^{(A)}]_{(a,s),(a',1-s)}=&[A_0]_{(a,s),(a',s)}.
\end{eqnarray}
Ancilla errors flip the ancilla with probability $p_{\mathrm{ancilla}}$, but these are dominated by $T_1$ decay, and so are highly asymmetric.
To account for this, we used different error rates $p_{\mathrm{anc},a,a'}$ for the four possible combinations of ancilla measurement at time $m-1$ and expected ancilla measurement at time $m$:
\begin{eqnarray}
[D_{\mathrm{anc},a,a'}^{(A)}]_{(a,s)(a',s)}&=-[A_0]_{(a,s),(a',s)},\\
{}[D_{\mathrm{anc},a,a'}^{(A)}]_{(a+1\;\mathrm{mod}\;2,s)(a',s)}&=-[A_0]_{(a,s),(a',s)}.
\end{eqnarray}
(Note that this asymmetry could not be accounted for in the data-qubit HMM as the state of the ancilla was not contained within the output vector.)
As with the data-qubit HMMs, we assume that ancilla leakage is HMM-state independent, as it is dominated by CZ gates during the time that the ancilla is either in $|+\rangle$ or $|-\rangle$.
We also assume that leakage (with rate $p_{\mathrm{leak}}$) and seepage (with rate $p_{\mathrm{seep}}$) have equal chances to flip the stabilizer state, as ancilla leakage has a good chance to cause additional error on the data qubits.

The ancilla-qubit HMMs need little adjustment between the $\SZZ$-check experiment and the experiment interleaving $\SZZ$ and $\SXX$ checks.
The $\HZZXX$-A HMM behaves almost identically to the $\HZZ$-A HMM, but we include in the state information on the $\SXX$ stabilizer as well as the $\SZZ$ stabilizer.
This leaves the states indexed as $(a,s_1,s_2)$.
The HMM needs to also keep track of which stabilizer is being measured. This may be achieved by shuffling the stabilizer labels at each timestep: for $a=0,1$, we set
\begin{equation}
[A_0]_{(a+s_1\;\mathrm{mod}\;2,s_2,s_1)(a,s_1,s_2)}=1.
\end{equation}
Other than this, the HMM follows the same equations as above (with the additional index added as expected.)

\begin{table*}
    \centering
    \begin{tabular}{|c|c|c|c|c|}
    \hline
    \hline
    \textbf{Error type} & $\bm{\HZZ}$\textbf{-A} & $\bm{\HZZ}$\textbf{-D} & $\bm{\HZZXX}$\textbf{-A} & $\bm{\HZZXX}$\textbf{-D} \\
    \hline
    \hline
    leakage [$p_{\mathrm{leak}}$] & $0.0040^*$ & 0.0064 & $0.0040^*$ & 0.0064 \\\hline
    seepage [$p_{\mathrm{seep}}$] & 0.101 & 0.108 & 0.101 & 0.103 \\\hline
    data-qubit error [$p_{\mathrm{data}}$] & 0.042 & 0.050 & 0.045 & 0.030 \\
    during leakage [$p_{\mathrm{data, leaked}}$] & - & 0.155 & - & 0.489 \\
    Y error (additional) [$p_{\mathrm{data, Y}}$] & - & - & - & 0.014 \\\hline
    readout error [$p_{\mathrm{readout}}$] & 0.011 & 0.004 & 0.027 & 0.014 \\\hline
    ancilla error [$p_{\mathrm{ancilla}}$] & 0.028 & 0.030 & - & 0.029 \\
    ($\MA[m-1]=1$, $\MA[m]=1$) [$p_{\mathrm{anc},0,0}$] & - & - & 0.001 & - \\
    ($\MA[m-1]=1$, $\MA[m]=-1$) [$p_{\mathrm{anc},0,1}$] & - & - & 0.021 & - \\
    ($\MA[m-1]=-1$, $\MA[m]=1$) [$p_{\mathrm{anc},1,0}$] & - & - & 0.044 & - \\
    ($\MA[m-1]=-1$, $\MA[m]=-1$) [$p_{\mathrm{anc},1,1}$] & - & - & 0.058 & - \\ 
    during leakage [$p_{\mathrm{ancilla, leaked}}$] & - & 0.113 & - & - \\\hline
    \end{tabular}
    \caption{Values of error rates used in the various HMMs in this work. All values are obtained by optimizing the likelihood of observing the given syndrome data except for the ancilla leakage rate (denoted $^*$) which is directly obtained from the experiments (as noted in the text).}
    \label{tab:HMM_params}
\end{table*}

\subsection{Uncertainty calculations}
All quoted uncertainties are an estimation of standard error of the mean (SEM).
SEMs for the independent device characterizations (\cref{sec:2.2}, \cref{tab:HMM_params}) are either obtained from at least three individually fitted repeated experiments ($T_2^{\mathrm{echo}}$, $T_1$, $T_2^{*}$, $\eta$, $e_{\mathrm{a}}$, $e_{\mathrm{a, ZZ}}$) or in the case that the quantitiy is only measured once ($e_{\mathrm{SQ}}$, $e_{\mathrm{CZ}}$, $L_{1}$), the SEM is estimated from least-squares fitting by the LmFit fitting module using the covariance matrix~\cite{LMFIT14}.

SEMs in the first-round Bell-state fidelities (\cref{fig:ZZ,fig:ZZXX}, \cref{sec:2.2,sec:2.5}) are obtained through bootstrapping. 
For bootstrapping, a data-set (in total 4096 runs with each 36 tomographic elements and 28 calibration points) is subdivided into four subsets and tomography is performed on each of these subsets individually.
As verification, subdivision was performed with eight subsets leading to similar SEMs.

SEMs in the multi-round experiment parameters (steady-state fidelities, decay constants) are also estimated from least-squares fitting by the LmFit fitting module using the covariance matrix~\cite{LMFIT14} (\cref{sec:2.3,sec:2.4,sec:2.5}).

\begin{acknowledgments}
We thank W.~Oliver and G.~Calusine for providing the parametric amplifier, J.~van~Oven and J.~de~Sterke for experimental assistance and F.~Battistel, C.~Beenakker,  C.~Eichler,  F.~Luthi,  B.~Terhal, and A.~Wallraff for discussions. 
\textbf{Funding:} This research is supported by the Office of the Director of National Intelligence (ODNI), Intelligence Advanced Research Projects Activity (IARPA), via the U.S. Army Research Office Grant No. W911NF-16-1-0071, and by Intel Corporation. T.E.O. is funded by Shell Global Solutions BV. The views and conclusions contained herein are those of the authors and should not be interpreted as necessarily representing the official policies or endorsements, either expressed or implied, of the ODNI, IARPA, or the
U.S. Government. X.F. was funded by China Scholarship Council (CSC).
\textbf{Author contributions:} C.C.B. performed the experiment. R.V. and M.W.B. designed the device with input from C.C.B. and B.T. N.M. and A.B. fabricated the device. T.E.O. devised the HMMs with input from B.T., B.V., and V.O. X.F. and M.A.R. contributed to the experimental setup and tune-up. C.C.B., T.E.O., and L.D.C. co-wrote the manuscript with feedback from all authors. L.D.C. supervised the project.
\textbf{Data and materials availability:} All data needed to evaluate the conclusions in the paper are present in the paper and/or the Supplementary Materials. Additional data related to this paper may be requested from the authors. 
\end{acknowledgments}


\begin{thebibliography}{52}%
\makeatletter
\providecommand \@ifxundefined [1]{%
 \@ifx{#1\undefined}
}%
\providecommand \@ifnum [1]{%
 \ifnum #1\expandafter \@firstoftwo
 \else \expandafter \@secondoftwo
 \fi
}%
\providecommand \@ifx [1]{%
 \ifx #1\expandafter \@firstoftwo
 \else \expandafter \@secondoftwo
 \fi
}%
\providecommand \natexlab [1]{#1}%
\providecommand \enquote  [1]{``#1''}%
\providecommand \bibnamefont  [1]{#1}%
\providecommand \bibfnamefont [1]{#1}%
\providecommand \citenamefont [1]{#1}%
\providecommand \href@noop [0]{\@secondoftwo}%
\providecommand \href [0]{\begingroup \@sanitize@url \@href}%
\providecommand \@href[1]{\@@startlink{#1}\@@href}%
\providecommand \@@href[1]{\endgroup#1\@@endlink}%
\providecommand \@sanitize@url [0]{\catcode `\\12\catcode `\$12\catcode
  `\&12\catcode `\#12\catcode `\^12\catcode `\_12\catcode `\%12\relax}%
\providecommand \@@startlink[1]{}%
\providecommand \@@endlink[0]{}%
\providecommand \url  [0]{\begingroup\@sanitize@url \@url }%
\providecommand \@url [1]{\endgroup\@href {#1}{\urlprefix }}%
\providecommand \urlprefix  [0]{URL }%
\providecommand \Eprint [0]{\href }%
\providecommand \doibase [0]{https://doi.org/}%
\providecommand \selectlanguage [0]{\@gobble}%
\providecommand \bibinfo  [0]{\@secondoftwo}%
\providecommand \bibfield  [0]{\@secondoftwo}%
\providecommand \translation [1]{[#1]}%
\providecommand \BibitemOpen [0]{}%
\providecommand \bibitemStop [0]{}%
\providecommand \bibitemNoStop [0]{.\EOS\space}%
\providecommand \EOS [0]{\spacefactor3000\relax}%
\providecommand \BibitemShut  [1]{\csname bibitem#1\endcsname}%
\let\auto@bib@innerbib\@empty
\bibitem [{\citenamefont {Terhal}(2015)}]{Terhal15}%
  \BibitemOpen
  \bibfield  {author} {\bibinfo {author} {\bibfnamefont {B.~M.}\ \bibnamefont
  {Terhal}},\ }\bibfield  {title} {\bibinfo {title} {Quantum error correction
  for quantum memories},\ }\href {https://doi.org/10.1103/RevModPhys.87.307}
  {\bibfield  {journal} {\bibinfo  {journal} {Rev. Mod. Phys.}\ }\textbf
  {\bibinfo {volume} {87}},\ \bibinfo {pages} {307} (\bibinfo {year}
  {2015})}\BibitemShut {NoStop}%
\bibitem [{\citenamefont {Barends}\ \emph {et~al.}(2014)\citenamefont
  {Barends}, \citenamefont {Kelly}, \citenamefont {Megrant}, \citenamefont
  {Veitia}, \citenamefont {Sank}, \citenamefont {Jeffrey}, \citenamefont
  {White}, \citenamefont {Mutus}, \citenamefont {Fowler}, \citenamefont
  {Campbell}, \citenamefont {Chen}, \citenamefont {Chen}, \citenamefont
  {Chiaro}, \citenamefont {Dunsworth}, \citenamefont {Neill}, \citenamefont
  {O'Malley}, \citenamefont {Roushan}, \citenamefont {Vainsencher},
  \citenamefont {Wenner}, \citenamefont {Korotkov}, \citenamefont {Cleland},\
  and\ \citenamefont {Martinis}}]{Barends14}%
  \BibitemOpen
  \bibfield  {author} {\bibinfo {author} {\bibfnamefont {R.}~\bibnamefont
  {Barends}}, \bibinfo {author} {\bibfnamefont {J.}~\bibnamefont {Kelly}},
  \bibinfo {author} {\bibfnamefont {A.}~\bibnamefont {Megrant}}, \bibinfo
  {author} {\bibfnamefont {A.}~\bibnamefont {Veitia}}, \bibinfo {author}
  {\bibfnamefont {D.}~\bibnamefont {Sank}}, \bibinfo {author} {\bibfnamefont
  {E.}~\bibnamefont {Jeffrey}}, \bibinfo {author} {\bibfnamefont {T.~C.}\
  \bibnamefont {White}}, \bibinfo {author} {\bibfnamefont {J.}~\bibnamefont
  {Mutus}}, \bibinfo {author} {\bibfnamefont {A.~G.}\ \bibnamefont {Fowler}},
  \bibinfo {author} {\bibfnamefont {B.}~\bibnamefont {Campbell}}, \bibinfo
  {author} {\bibfnamefont {Y.}~\bibnamefont {Chen}}, \bibinfo {author}
  {\bibfnamefont {Z.}~\bibnamefont {Chen}}, \bibinfo {author} {\bibfnamefont
  {B.}~\bibnamefont {Chiaro}}, \bibinfo {author} {\bibfnamefont
  {A.}~\bibnamefont {Dunsworth}}, \bibinfo {author} {\bibfnamefont
  {C.}~\bibnamefont {Neill}}, \bibinfo {author} {\bibfnamefont
  {P.}~\bibnamefont {O'Malley}}, \bibinfo {author} {\bibfnamefont
  {P.}~\bibnamefont {Roushan}}, \bibinfo {author} {\bibfnamefont
  {A.}~\bibnamefont {Vainsencher}}, \bibinfo {author} {\bibfnamefont
  {J.}~\bibnamefont {Wenner}}, \bibinfo {author} {\bibfnamefont {A.~N.}\
  \bibnamefont {Korotkov}}, \bibinfo {author} {\bibfnamefont {A.~N.}\
  \bibnamefont {Cleland}},\ and\ \bibinfo {author} {\bibfnamefont {J.~M.}\
  \bibnamefont {Martinis}},\ }\bibfield  {title} {\bibinfo {title}
  {{Superconducting quantum circuits at the surface code threshold for fault
  tolerance}},\ }\href
  {http://www.nature.com/nature/journal/v508/n7497/abs/nature13171.html}
  {\bibfield  {journal} {\bibinfo  {journal} {Nature}\ }\textbf {\bibinfo
  {volume} {508}},\ \bibinfo {pages} {500} (\bibinfo {year}
  {2014})}\BibitemShut {NoStop}%
\bibitem [{\citenamefont {Harty}\ \emph {et~al.}(2014)\citenamefont {Harty},
  \citenamefont {Allcock}, \citenamefont {Ballance}, \citenamefont {Guidoni},
  \citenamefont {Janacek}, \citenamefont {Linke}, \citenamefont {Stacey},\ and\
  \citenamefont {Lucas}}]{Harty14}%
  \BibitemOpen
  \bibfield  {author} {\bibinfo {author} {\bibfnamefont {T.~P.}\ \bibnamefont
  {Harty}}, \bibinfo {author} {\bibfnamefont {D.~T.~C.}\ \bibnamefont
  {Allcock}}, \bibinfo {author} {\bibfnamefont {C.~J.}\ \bibnamefont
  {Ballance}}, \bibinfo {author} {\bibfnamefont {L.}~\bibnamefont {Guidoni}},
  \bibinfo {author} {\bibfnamefont {H.~A.}\ \bibnamefont {Janacek}}, \bibinfo
  {author} {\bibfnamefont {N.~M.}\ \bibnamefont {Linke}}, \bibinfo {author}
  {\bibfnamefont {D.~N.}\ \bibnamefont {Stacey}},\ and\ \bibinfo {author}
  {\bibfnamefont {D.~M.}\ \bibnamefont {Lucas}},\ }\bibfield  {title} {\bibinfo
  {title} {High-fidelity preparation, gates, memory, and readout of a
  trapped-ion quantum bit},\ }\href
  {https://doi.org/10.1103/PhysRevLett.113.220501} {\bibfield  {journal}
  {\bibinfo  {journal} {Phys. Rev. Lett.}\ }\textbf {\bibinfo {volume} {113}},\
  \bibinfo {pages} {220501} (\bibinfo {year} {2014})}\BibitemShut {NoStop}%
\bibitem [{\citenamefont {Ballance}\ \emph {et~al.}(2016)\citenamefont
  {Ballance}, \citenamefont {Harty}, \citenamefont {Linke}, \citenamefont
  {Sepiol},\ and\ \citenamefont {Lucas}}]{Ballance16}%
  \BibitemOpen
  \bibfield  {author} {\bibinfo {author} {\bibfnamefont {C.~J.}\ \bibnamefont
  {Ballance}}, \bibinfo {author} {\bibfnamefont {T.~P.}\ \bibnamefont {Harty}},
  \bibinfo {author} {\bibfnamefont {N.~M.}\ \bibnamefont {Linke}}, \bibinfo
  {author} {\bibfnamefont {M.~A.}\ \bibnamefont {Sepiol}},\ and\ \bibinfo
  {author} {\bibfnamefont {D.~M.}\ \bibnamefont {Lucas}},\ }\bibfield  {title}
  {\bibinfo {title} {High-fidelity quantum logic gates using trapped-ion
  hyperfine qubits},\ }\href {https://doi.org/10.1103/PhysRevLett.117.060504}
  {\bibfield  {journal} {\bibinfo  {journal} {Phys. Rev. Lett.}\ }\textbf
  {\bibinfo {volume} {117}},\ \bibinfo {pages} {060504} (\bibinfo {year}
  {2016})}\BibitemShut {NoStop}%
\bibitem [{\citenamefont {Rol}\ \emph {et~al.}(2019)\citenamefont {Rol},
  \citenamefont {Battistel}, \citenamefont {Malinowski}, \citenamefont
  {Bultink}, \citenamefont {Tarasinski}, \citenamefont {Vollmer}, \citenamefont
  {Haider}, \citenamefont {Muthusubramanian}, \citenamefont {Bruno},
  \citenamefont {Terhal},\ and\ \citenamefont {DiCarlo}}]{Rol19a}%
  \BibitemOpen
  \bibfield  {author} {\bibinfo {author} {\bibfnamefont {M.~A.}\ \bibnamefont
  {Rol}}, \bibinfo {author} {\bibfnamefont {F.}~\bibnamefont {Battistel}},
  \bibinfo {author} {\bibfnamefont {F.~K.}\ \bibnamefont {Malinowski}},
  \bibinfo {author} {\bibfnamefont {C.~C.}\ \bibnamefont {Bultink}}, \bibinfo
  {author} {\bibfnamefont {B.~M.}\ \bibnamefont {Tarasinski}}, \bibinfo
  {author} {\bibfnamefont {R.}~\bibnamefont {Vollmer}}, \bibinfo {author}
  {\bibfnamefont {N.}~\bibnamefont {Haider}}, \bibinfo {author} {\bibfnamefont
  {N.}~\bibnamefont {Muthusubramanian}}, \bibinfo {author} {\bibfnamefont
  {A.}~\bibnamefont {Bruno}}, \bibinfo {author} {\bibfnamefont {B.~M.}\
  \bibnamefont {Terhal}},\ and\ \bibinfo {author} {\bibfnamefont
  {L.}~\bibnamefont {DiCarlo}},\ }\bibfield  {title} {\bibinfo {title} {Fast,
  high-fidelity conditional-phase gate exploiting leakage interference in
  weakly anharmonic superconducting qubits},\ }\href
  {https://doi.org/10.1103/PhysRevLett.123.120502} {\bibfield  {journal}
  {\bibinfo  {journal} {Phys. Rev. Lett.}\ }\textbf {\bibinfo {volume} {123}},\
  \bibinfo {pages} {120502} (\bibinfo {year} {2019})}\BibitemShut {NoStop}%
\bibitem [{\citenamefont {Jeffrey}\ \emph {et~al.}(2014)\citenamefont
  {Jeffrey}, \citenamefont {Sank}, \citenamefont {Mutus}, \citenamefont
  {White}, \citenamefont {Kelly}, \citenamefont {Barends}, \citenamefont
  {Chen}, \citenamefont {Chen}, \citenamefont {Chiaro}, \citenamefont
  {Dunsworth}, \citenamefont {Megrant}, \citenamefont {O'Malley}, \citenamefont
  {Neill}, \citenamefont {Roushan}, \citenamefont {Vainsencher}, \citenamefont
  {Wenner}, \citenamefont {Cleland},\ and\ \citenamefont
  {Martinis}}]{Jeffrey14}%
  \BibitemOpen
  \bibfield  {author} {\bibinfo {author} {\bibfnamefont {E.}~\bibnamefont
  {Jeffrey}}, \bibinfo {author} {\bibfnamefont {D.}~\bibnamefont {Sank}},
  \bibinfo {author} {\bibfnamefont {J.~Y.}\ \bibnamefont {Mutus}}, \bibinfo
  {author} {\bibfnamefont {T.~C.}\ \bibnamefont {White}}, \bibinfo {author}
  {\bibfnamefont {J.}~\bibnamefont {Kelly}}, \bibinfo {author} {\bibfnamefont
  {R.}~\bibnamefont {Barends}}, \bibinfo {author} {\bibfnamefont
  {Y.}~\bibnamefont {Chen}}, \bibinfo {author} {\bibfnamefont {Z.}~\bibnamefont
  {Chen}}, \bibinfo {author} {\bibfnamefont {B.}~\bibnamefont {Chiaro}},
  \bibinfo {author} {\bibfnamefont {A.}~\bibnamefont {Dunsworth}}, \bibinfo
  {author} {\bibfnamefont {A.}~\bibnamefont {Megrant}}, \bibinfo {author}
  {\bibfnamefont {P.~J.~J.}\ \bibnamefont {O'Malley}}, \bibinfo {author}
  {\bibfnamefont {C.}~\bibnamefont {Neill}}, \bibinfo {author} {\bibfnamefont
  {P.}~\bibnamefont {Roushan}}, \bibinfo {author} {\bibfnamefont
  {A.}~\bibnamefont {Vainsencher}}, \bibinfo {author} {\bibfnamefont
  {J.}~\bibnamefont {Wenner}}, \bibinfo {author} {\bibfnamefont {A.~N.}\
  \bibnamefont {Cleland}},\ and\ \bibinfo {author} {\bibfnamefont {J.~M.}\
  \bibnamefont {Martinis}},\ }\bibfield  {title} {\bibinfo {title} {Fast
  accurate state measurement with superconducting qubits},\ }\href
  {https://journals.aps.org/prl/abstract/10.1103/PhysRevLett.112.190504}
  {\bibfield  {journal} {\bibinfo  {journal} {Phys. Rev. Lett.}\ }\textbf
  {\bibinfo {volume} {112}},\ \bibinfo {pages} {190504} (\bibinfo {year}
  {2014})}\BibitemShut {NoStop}%
\bibitem [{\citenamefont {Bultink}\ \emph {et~al.}(2016)\citenamefont
  {Bultink}, \citenamefont {Rol}, \citenamefont {O'Brien}, \citenamefont {Fu},
  \citenamefont {Dikken}, \citenamefont {Dickel}, \citenamefont {Vermeulen},
  \citenamefont {de~Sterke}, \citenamefont {Bruno}, \citenamefont {Schouten},\
  and\ \citenamefont {DiCarlo}}]{Bultink16}%
  \BibitemOpen
  \bibfield  {author} {\bibinfo {author} {\bibfnamefont {C.~C.}\ \bibnamefont
  {Bultink}}, \bibinfo {author} {\bibfnamefont {M.~A.}\ \bibnamefont {Rol}},
  \bibinfo {author} {\bibfnamefont {T.~E.}\ \bibnamefont {O'Brien}}, \bibinfo
  {author} {\bibfnamefont {X.}~\bibnamefont {Fu}}, \bibinfo {author}
  {\bibfnamefont {B.~C.~S.}\ \bibnamefont {Dikken}}, \bibinfo {author}
  {\bibfnamefont {C.}~\bibnamefont {Dickel}}, \bibinfo {author} {\bibfnamefont
  {R.~F.~L.}\ \bibnamefont {Vermeulen}}, \bibinfo {author} {\bibfnamefont
  {J.~C.}\ \bibnamefont {de~Sterke}}, \bibinfo {author} {\bibfnamefont
  {A.}~\bibnamefont {Bruno}}, \bibinfo {author} {\bibfnamefont {R.~N.}\
  \bibnamefont {Schouten}},\ and\ \bibinfo {author} {\bibfnamefont
  {L.}~\bibnamefont {DiCarlo}},\ }\bibfield  {title} {\bibinfo {title} {Active
  resonator reset in the nonlinear dispersive regime of circuit {Q}{E}{D}},\
  }\href {https://link.aps.org/doi/10.1103/PhysRevApplied.6.034008} {\bibfield
  {journal} {\bibinfo  {journal} {Phys. Rev. Appl.}\ }\textbf {\bibinfo
  {volume} {6}},\ \bibinfo {pages} {034008} (\bibinfo {year}
  {2016})}\BibitemShut {NoStop}%
\bibitem [{\citenamefont {Heinsoo}\ \emph {et~al.}(2018)\citenamefont
  {Heinsoo}, \citenamefont {Andersen}, \citenamefont {Remm}, \citenamefont
  {Krinner}, \citenamefont {Walter}, \citenamefont {Salath\'e}, \citenamefont
  {Gasparinetti}, \citenamefont {Besse}, \citenamefont
  {Poto\ifmmode~\check{c}\else \v{c}\fi{}nik}, \citenamefont {Wallraff},\ and\
  \citenamefont {Eichler}}]{Heinsoo18}%
  \BibitemOpen
  \bibfield  {author} {\bibinfo {author} {\bibfnamefont {J.}~\bibnamefont
  {Heinsoo}}, \bibinfo {author} {\bibfnamefont {C.~K.}\ \bibnamefont
  {Andersen}}, \bibinfo {author} {\bibfnamefont {A.}~\bibnamefont {Remm}},
  \bibinfo {author} {\bibfnamefont {S.}~\bibnamefont {Krinner}}, \bibinfo
  {author} {\bibfnamefont {T.}~\bibnamefont {Walter}}, \bibinfo {author}
  {\bibfnamefont {Y.}~\bibnamefont {Salath\'e}}, \bibinfo {author}
  {\bibfnamefont {S.}~\bibnamefont {Gasparinetti}}, \bibinfo {author}
  {\bibfnamefont {J.-C.}\ \bibnamefont {Besse}}, \bibinfo {author}
  {\bibfnamefont {A.}~\bibnamefont {Poto\ifmmode~\check{c}\else
  \v{c}\fi{}nik}}, \bibinfo {author} {\bibfnamefont {A.}~\bibnamefont
  {Wallraff}},\ and\ \bibinfo {author} {\bibfnamefont {C.}~\bibnamefont
  {Eichler}},\ }\bibfield  {title} {\bibinfo {title} {Rapid high-fidelity
  multiplexed readout of superconducting qubits},\ }\href
  {https://doi.org/10.1103/PhysRevApplied.10.034040} {\bibfield  {journal}
  {\bibinfo  {journal} {Phys. Rev. Appl.}\ }\textbf {\bibinfo {volume} {10}},\
  \bibinfo {pages} {034040} (\bibinfo {year} {2018})}\BibitemShut {NoStop}%
\bibitem [{\citenamefont {Raussendorf}\ and\ \citenamefont
  {Harrington}(2007)}]{Raussendorf07}%
  \BibitemOpen
  \bibfield  {author} {\bibinfo {author} {\bibfnamefont {R.}~\bibnamefont
  {Raussendorf}}\ and\ \bibinfo {author} {\bibfnamefont {J.}~\bibnamefont
  {Harrington}},\ }\bibfield  {title} {\bibinfo {title} {Fault-tolerant quantum
  computation with high threshold in two dimensions},\ }\href
  {https://doi.org/10.1103/PhysRevLett.98.190504} {\bibfield  {journal}
  {\bibinfo  {journal} {Phys. Rev. Lett.}\ }\textbf {\bibinfo {volume} {98}},\
  \bibinfo {pages} {190504} (\bibinfo {year} {2007})}\BibitemShut {NoStop}%
\bibitem [{\citenamefont {Fowler}\ \emph {et~al.}(2012)\citenamefont {Fowler},
  \citenamefont {Mariantoni}, \citenamefont {Martinis},\ and\ \citenamefont
  {Cleland}}]{Fowler12}%
  \BibitemOpen
  \bibfield  {author} {\bibinfo {author} {\bibfnamefont {A.~G.}\ \bibnamefont
  {Fowler}}, \bibinfo {author} {\bibfnamefont {M.}~\bibnamefont {Mariantoni}},
  \bibinfo {author} {\bibfnamefont {J.~M.}\ \bibnamefont {Martinis}},\ and\
  \bibinfo {author} {\bibfnamefont {A.~N.}\ \bibnamefont {Cleland}},\
  }\bibfield  {title} {\bibinfo {title} {Surface codes: Towards practical
  large-scale quantum computation},\ }\href
  {https://link.aps.org/doi/10.1103/PhysRevA.86.032324} {\bibfield  {journal}
  {\bibinfo  {journal} {Phys. Rev. A}\ }\textbf {\bibinfo {volume} {86}},\
  \bibinfo {pages} {032324} (\bibinfo {year} {2012})}\BibitemShut {NoStop}%
\bibitem [{\citenamefont {Bombin}\ and\ \citenamefont
  {Martin-Delgado}(2007)}]{Bombin07}%
  \BibitemOpen
  \bibfield  {author} {\bibinfo {author} {\bibfnamefont {H.}~\bibnamefont
  {Bombin}}\ and\ \bibinfo {author} {\bibfnamefont {M.~A.}\ \bibnamefont
  {Martin-Delgado}},\ }\bibfield  {title} {\bibinfo {title} {Topological
  computation without braiding},\ }\href
  {https://doi.org/10.1103/PhysRevLett.98.160502} {\bibfield  {journal}
  {\bibinfo  {journal} {Phys. Rev. Lett.}\ }\textbf {\bibinfo {volume} {98}},\
  \bibinfo {pages} {160502} (\bibinfo {year} {2007})}\BibitemShut {NoStop}%
\bibitem [{\citenamefont {Brown}\ and\ \citenamefont {Brown}(2018)}]{Brown18}%
  \BibitemOpen
  \bibfield  {author} {\bibinfo {author} {\bibfnamefont {N.~C.}\ \bibnamefont
  {Brown}}\ and\ \bibinfo {author} {\bibfnamefont {K.~R.}\ \bibnamefont
  {Brown}},\ }\bibfield  {title} {\bibinfo {title} {Comparing zeeman qubits to
  hyperfine qubits in the context of the surface code: $^{174}\mathrm{Yb}^{+}$
  and $^{171}\mathrm{Yb}^{+}$},\ }\href
  {https://doi.org/10.1103/PhysRevA.97.052301} {\bibfield  {journal} {\bibinfo
  {journal} {Phys. Rev. A}\ }\textbf {\bibinfo {volume} {97}},\ \bibinfo
  {pages} {052301} (\bibinfo {year} {2018})}\BibitemShut {NoStop}%
\bibitem [{\citenamefont {Andrews}\ \emph {et~al.}(2019)\citenamefont
  {Andrews}, \citenamefont {Jones}, \citenamefont {Reed}, \citenamefont
  {Jones}, \citenamefont {Ha}, \citenamefont {Jura}, \citenamefont {Kerckhoff},
  \citenamefont {Levendorf}, \citenamefont {Meenehan}, \citenamefont {Merkel},
  \citenamefont {Smith}, \citenamefont {Sun}, \citenamefont {Weinstein},
  \citenamefont {Rakher}, \citenamefont {Ladd},\ and\ \citenamefont
  {Borselli}}]{Andrews19}%
  \BibitemOpen
  \bibfield  {author} {\bibinfo {author} {\bibfnamefont {R.~W.}\ \bibnamefont
  {Andrews}}, \bibinfo {author} {\bibfnamefont {C.}~\bibnamefont {Jones}},
  \bibinfo {author} {\bibfnamefont {M.~D.}\ \bibnamefont {Reed}}, \bibinfo
  {author} {\bibfnamefont {A.~M.}\ \bibnamefont {Jones}}, \bibinfo {author}
  {\bibfnamefont {S.~D.}\ \bibnamefont {Ha}}, \bibinfo {author} {\bibfnamefont
  {M.~P.}\ \bibnamefont {Jura}}, \bibinfo {author} {\bibfnamefont
  {J.}~\bibnamefont {Kerckhoff}}, \bibinfo {author} {\bibfnamefont
  {M.}~\bibnamefont {Levendorf}}, \bibinfo {author} {\bibfnamefont
  {S.}~\bibnamefont {Meenehan}}, \bibinfo {author} {\bibfnamefont {S.~T.}\
  \bibnamefont {Merkel}}, \bibinfo {author} {\bibfnamefont {A.}~\bibnamefont
  {Smith}}, \bibinfo {author} {\bibfnamefont {B.}~\bibnamefont {Sun}}, \bibinfo
  {author} {\bibfnamefont {A.~J.}\ \bibnamefont {Weinstein}}, \bibinfo {author}
  {\bibfnamefont {M.~T.}\ \bibnamefont {Rakher}}, \bibinfo {author}
  {\bibfnamefont {T.~D.}\ \bibnamefont {Ladd}},\ and\ \bibinfo {author}
  {\bibfnamefont {M.~G.}\ \bibnamefont {Borselli}},\ }\bibfield  {title}
  {\bibinfo {title} {{Quantifying error and leakage in an encoded Si/SiGe
  triple-dot qubit}},\ }\href {https://doi.org/10.1038/s41565-019-0500-4}
  {\bibfield  {journal} {\bibinfo  {journal} {Nat.\ Nanotechnol.}\ }\textbf
  {\bibinfo {volume} {14}},\ \bibinfo {pages} {747} (\bibinfo {year}
  {2019})}\BibitemShut {NoStop}%
\bibitem [{\citenamefont {Aliferis}\ and\ \citenamefont
  {Terhal}(2007)}]{Aliferis07}%
  \BibitemOpen
  \bibfield  {author} {\bibinfo {author} {\bibfnamefont {P.}~\bibnamefont
  {Aliferis}}\ and\ \bibinfo {author} {\bibfnamefont {B.~M.}\ \bibnamefont
  {Terhal}},\ }\bibfield  {title} {\bibinfo {title} {Fault-tolerant quantum
  computation for local leakage faults},\ }\href
  {http://dl.acm.org/citation.cfm?id=2011706.2011715} {\bibfield  {journal}
  {\bibinfo  {journal} {Quantum Info. Comput.}\ }\textbf {\bibinfo {volume}
  {7}},\ \bibinfo {pages} {139} (\bibinfo {year} {2007})}\BibitemShut {NoStop}%
\bibitem [{\citenamefont {Fowler}(2013)}]{Fowler13}%
  \BibitemOpen
  \bibfield  {author} {\bibinfo {author} {\bibfnamefont {A.~G.}\ \bibnamefont
  {Fowler}},\ }\bibfield  {title} {\bibinfo {title} {Coping with qubit leakage
  in topological codes},\ }\href
  {https://link.aps.org/doi/10.1103/PhysRevA.88.042308} {\bibfield  {journal}
  {\bibinfo  {journal} {Phys. Rev. A}\ }\textbf {\bibinfo {volume} {88}},\
  \bibinfo {pages} {042308} (\bibinfo {year} {2013})}\BibitemShut {NoStop}%
\bibitem [{\citenamefont {Ghosh}\ and\ \citenamefont {Fowler}(2015)}]{Ghosh15}%
  \BibitemOpen
  \bibfield  {author} {\bibinfo {author} {\bibfnamefont {J.}~\bibnamefont
  {Ghosh}}\ and\ \bibinfo {author} {\bibfnamefont {A.~G.}\ \bibnamefont
  {Fowler}},\ }\bibfield  {title} {\bibinfo {title} {Leakage-resilient approach
  to fault-tolerant quantum computing with superconducting elements},\ }\href
  {https://doi.org/10.1103/PhysRevA.91.020302} {\bibfield  {journal} {\bibinfo
  {journal} {Phys. Rev. A}\ }\textbf {\bibinfo {volume} {91}},\ \bibinfo
  {pages} {020302} (\bibinfo {year} {2015})}\BibitemShut {NoStop}%
\bibitem [{\citenamefont {Suchara}\ \emph {et~al.}(2015)\citenamefont
  {Suchara}, \citenamefont {Cross},\ and\ \citenamefont
  {Gambetta}}]{Suchara15}%
  \BibitemOpen
  \bibfield  {author} {\bibinfo {author} {\bibfnamefont {M.}~\bibnamefont
  {Suchara}}, \bibinfo {author} {\bibfnamefont {A.~W.}\ \bibnamefont {Cross}},\
  and\ \bibinfo {author} {\bibfnamefont {J.~M.}\ \bibnamefont {Gambetta}},\
  }\bibfield  {title} {\bibinfo {title} {Leakage suppression in the toric
  code},\ }\href {http://dl.acm.org/citation.cfm?id=2871350.2871358} {\bibfield
   {journal} {\bibinfo  {journal} {Quantum Info. Comput.}\ }\textbf {\bibinfo
  {volume} {15}},\ \bibinfo {pages} {997} (\bibinfo {year} {2015})}\BibitemShut
  {NoStop}%
\bibitem [{\citenamefont {Brown}\ \emph {et~al.}(2019)\citenamefont {Brown},
  \citenamefont {Newman},\ and\ \citenamefont {Brown}}]{Brown19}%
  \BibitemOpen
  \bibfield  {author} {\bibinfo {author} {\bibfnamefont {N.~C.}\ \bibnamefont
  {Brown}}, \bibinfo {author} {\bibfnamefont {M.}~\bibnamefont {Newman}},\ and\
  \bibinfo {author} {\bibfnamefont {K.~R.}\ \bibnamefont {Brown}},\ }\bibfield
  {title} {\bibinfo {title} {Handling leakage with subsystem codes},\ }\href
  {https://doi.org/10.1088/1367-2630/ab3372} {\bibfield  {journal} {\bibinfo
  {journal} {New Journal of Physics}\ }\textbf {\bibinfo {volume} {21}},\
  \bibinfo {pages} {073055} (\bibinfo {year} {2019})}\BibitemShut {NoStop}%
\bibitem [{\citenamefont {Cai}\ \emph {et~al.}(2019)\citenamefont {Cai},
  \citenamefont {Fogarty}, \citenamefont {Schaal}, \citenamefont
  {Patom{\"{a}}ki}, \citenamefont {Benjamin},\ and\ \citenamefont
  {Morton}}]{Cai19}%
  \BibitemOpen
  \bibfield  {author} {\bibinfo {author} {\bibfnamefont {Z.}~\bibnamefont
  {Cai}}, \bibinfo {author} {\bibfnamefont {M.~A.}\ \bibnamefont {Fogarty}},
  \bibinfo {author} {\bibfnamefont {S.}~\bibnamefont {Schaal}}, \bibinfo
  {author} {\bibfnamefont {S.}~\bibnamefont {Patom{\"{a}}ki}}, \bibinfo
  {author} {\bibfnamefont {S.~C.}\ \bibnamefont {Benjamin}},\ and\ \bibinfo
  {author} {\bibfnamefont {J.~J.~L.}\ \bibnamefont {Morton}},\ }\bibfield
  {title} {\bibinfo {title} {A {S}ilicon {S}urface {C}ode {A}rchitecture
  {R}esilient {A}gainst {L}eakage {E}rrors},\ }\href
  {https://doi.org/10.22331/q-2019-12-09-212} {\bibfield  {journal} {\bibinfo
  {journal} {{Quantum}}\ }\textbf {\bibinfo {volume} {3}},\ \bibinfo {pages}
  {212} (\bibinfo {year} {2019})}\BibitemShut {NoStop}%
\bibitem [{\citenamefont {Rist\`{e}}\ \emph {et~al.}(2013)\citenamefont
  {Rist\`{e}}, \citenamefont {Dukalski}, \citenamefont {Watson}, \citenamefont
  {de~Lange}, \citenamefont {Tiggelman}, \citenamefont {Blanter}, \citenamefont
  {Lehnert}, \citenamefont {Schouten},\ and\ \citenamefont
  {DiCarlo}}]{Riste13b}%
  \BibitemOpen
  \bibfield  {author} {\bibinfo {author} {\bibfnamefont {D.}~\bibnamefont
  {Rist\`{e}}}, \bibinfo {author} {\bibfnamefont {M.}~\bibnamefont {Dukalski}},
  \bibinfo {author} {\bibfnamefont {C.~A.}\ \bibnamefont {Watson}}, \bibinfo
  {author} {\bibfnamefont {G.}~\bibnamefont {de~Lange}}, \bibinfo {author}
  {\bibfnamefont {M.~J.}\ \bibnamefont {Tiggelman}}, \bibinfo {author}
  {\bibfnamefont {Y.~M.}\ \bibnamefont {Blanter}}, \bibinfo {author}
  {\bibfnamefont {K.~W.}\ \bibnamefont {Lehnert}}, \bibinfo {author}
  {\bibfnamefont {R.~N.}\ \bibnamefont {Schouten}},\ and\ \bibinfo {author}
  {\bibfnamefont {L.}~\bibnamefont {DiCarlo}},\ }\bibfield  {title} {\bibinfo
  {title} {{Deterministic entanglement of superconducting qubits by parity
  measurement and feedback}},\ }\href {http://dx.doi.org/10.1038/nature12513}
  {\bibfield  {journal} {\bibinfo  {journal} {Nature}\ }\textbf {\bibinfo
  {volume} {502}},\ \bibinfo {pages} {350} (\bibinfo {year}
  {2013})}\BibitemShut {NoStop}%
\bibitem [{\citenamefont {Liu}\ \emph {et~al.}(2016)\citenamefont {Liu},
  \citenamefont {Shankar}, \citenamefont {Ofek}, \citenamefont {Hatridge},
  \citenamefont {Narla}, \citenamefont {Sliwa}, \citenamefont {Frunzio},
  \citenamefont {Schoelkopf},\ and\ \citenamefont {Devoret}}]{Liu16}%
  \BibitemOpen
  \bibfield  {author} {\bibinfo {author} {\bibfnamefont {Y.}~\bibnamefont
  {Liu}}, \bibinfo {author} {\bibfnamefont {S.}~\bibnamefont {Shankar}},
  \bibinfo {author} {\bibfnamefont {N.}~\bibnamefont {Ofek}}, \bibinfo {author}
  {\bibfnamefont {M.}~\bibnamefont {Hatridge}}, \bibinfo {author}
  {\bibfnamefont {A.}~\bibnamefont {Narla}}, \bibinfo {author} {\bibfnamefont
  {K.~M.}\ \bibnamefont {Sliwa}}, \bibinfo {author} {\bibfnamefont
  {L.}~\bibnamefont {Frunzio}}, \bibinfo {author} {\bibfnamefont {R.~J.}\
  \bibnamefont {Schoelkopf}},\ and\ \bibinfo {author} {\bibfnamefont {M.~H.}\
  \bibnamefont {Devoret}},\ }\bibfield  {title} {\bibinfo {title} {Comparing
  and combining measurement-based and driven-dissipative entanglement
  stabilization},\ }\href {https://doi.org/10.1103/PhysRevX.6.011022}
  {\bibfield  {journal} {\bibinfo  {journal} {Phys. Rev. X}\ }\textbf {\bibinfo
  {volume} {6}},\ \bibinfo {pages} {011022} (\bibinfo {year}
  {2016})}\BibitemShut {NoStop}%
\bibitem [{\citenamefont {C\'orcoles}\ \emph {et~al.}(2015)\citenamefont
  {C\'orcoles}, \citenamefont {Magesan}, \citenamefont {Srinivasan},
  \citenamefont {Cross}, \citenamefont {Steffen}, \citenamefont {Gambetta},\
  and\ \citenamefont {Chow}}]{Corcoles15}%
  \BibitemOpen
  \bibfield  {author} {\bibinfo {author} {\bibfnamefont {A.~D.}\ \bibnamefont
  {C\'orcoles}}, \bibinfo {author} {\bibfnamefont {E.}~\bibnamefont {Magesan}},
  \bibinfo {author} {\bibfnamefont {S.~J.}\ \bibnamefont {Srinivasan}},
  \bibinfo {author} {\bibfnamefont {A.~W.}\ \bibnamefont {Cross}}, \bibinfo
  {author} {\bibfnamefont {M.}~\bibnamefont {Steffen}}, \bibinfo {author}
  {\bibfnamefont {J.~M.}\ \bibnamefont {Gambetta}},\ and\ \bibinfo {author}
  {\bibfnamefont {J.~M.}\ \bibnamefont {Chow}},\ }\bibfield  {title} {\bibinfo
  {title} {{Demonstration of a quantum error detection code using a square
  lattice of four superconducting qubits}},\ }\href
  {https://www.nature.com/articles/ncomms7979} {\bibfield  {journal} {\bibinfo
  {journal} {Nat.\ Commun.}\ }\textbf {\bibinfo {volume} {{6}}},\ \bibinfo
  {pages} {6979} (\bibinfo {year} {{2015}})}\BibitemShut {NoStop}%
\bibitem [{\citenamefont {Rist\`{e}}\ \emph {et~al.}(2015)\citenamefont
  {Rist\`{e}}, \citenamefont {Poletto}, \citenamefont {Huang}, \citenamefont
  {Bruno}, \citenamefont {Vesterinen}, \citenamefont {Saira},\ and\
  \citenamefont {DiCarlo}}]{Riste15}%
  \BibitemOpen
  \bibfield  {author} {\bibinfo {author} {\bibfnamefont {D.}~\bibnamefont
  {Rist\`{e}}}, \bibinfo {author} {\bibfnamefont {S.}~\bibnamefont {Poletto}},
  \bibinfo {author} {\bibfnamefont {M.~Z.}\ \bibnamefont {Huang}}, \bibinfo
  {author} {\bibfnamefont {A.}~\bibnamefont {Bruno}}, \bibinfo {author}
  {\bibfnamefont {V.}~\bibnamefont {Vesterinen}}, \bibinfo {author}
  {\bibfnamefont {O.~P.}\ \bibnamefont {Saira}},\ and\ \bibinfo {author}
  {\bibfnamefont {L.}~\bibnamefont {DiCarlo}},\ }\bibfield  {title} {\bibinfo
  {title} {{Detecting bit-flip errors in a logical qubit using stabilizer
  measurements}},\ }\href {https://www.nature.com/articles/ncomms7983}
  {\bibfield  {journal} {\bibinfo  {journal} {Nat.\ Commun.}\ }\textbf
  {\bibinfo {volume} {{6}}},\ \bibinfo {pages} {6983} (\bibinfo {year}
  {{2015}})}\BibitemShut {NoStop}%
\bibitem [{\citenamefont {Kelly}\ \emph {et~al.}(2015)\citenamefont {Kelly},
  \citenamefont {Barends}, \citenamefont {Fowler}, \citenamefont {Megrant},
  \citenamefont {Jeffrey}, \citenamefont {White}, \citenamefont {Sank},
  \citenamefont {Mutus}, \citenamefont {Campbell}, \citenamefont {Chen} \emph
  {et~al.}}]{Kelly15}%
  \BibitemOpen
  \bibfield  {author} {\bibinfo {author} {\bibfnamefont {J.}~\bibnamefont
  {Kelly}}, \bibinfo {author} {\bibfnamefont {R.}~\bibnamefont {Barends}},
  \bibinfo {author} {\bibfnamefont {A.}~\bibnamefont {Fowler}}, \bibinfo
  {author} {\bibfnamefont {A.}~\bibnamefont {Megrant}}, \bibinfo {author}
  {\bibfnamefont {E.}~\bibnamefont {Jeffrey}}, \bibinfo {author} {\bibfnamefont
  {T.}~\bibnamefont {White}}, \bibinfo {author} {\bibfnamefont
  {D.}~\bibnamefont {Sank}}, \bibinfo {author} {\bibfnamefont {J.}~\bibnamefont
  {Mutus}}, \bibinfo {author} {\bibfnamefont {B.}~\bibnamefont {Campbell}},
  \bibinfo {author} {\bibfnamefont {Y.}~\bibnamefont {Chen}}, \emph {et~al.},\
  }\bibfield  {title} {\bibinfo {title} {State preservation by repetitive error
  detection in a superconducting quantum circuit},\ }\href
  {https://www.nature.com/nature/journal/v519/n7541/full/nature14270.html}
  {\bibfield  {journal} {\bibinfo  {journal} {Nature}\ }\textbf {\bibinfo
  {volume} {519}},\ \bibinfo {pages} {66} (\bibinfo {year} {2015})}\BibitemShut
  {NoStop}%
\bibitem [{\citenamefont {Takita}\ \emph {et~al.}(2016)\citenamefont {Takita},
  \citenamefont {C\'orcoles}, \citenamefont {Magesan}, \citenamefont {Abdo},
  \citenamefont {Brink}, \citenamefont {Cross}, \citenamefont {Chow},\ and\
  \citenamefont {Gambetta}}]{Takita16}%
  \BibitemOpen
  \bibfield  {author} {\bibinfo {author} {\bibfnamefont {M.}~\bibnamefont
  {Takita}}, \bibinfo {author} {\bibfnamefont {A.~D.}\ \bibnamefont
  {C\'orcoles}}, \bibinfo {author} {\bibfnamefont {E.}~\bibnamefont {Magesan}},
  \bibinfo {author} {\bibfnamefont {B.}~\bibnamefont {Abdo}}, \bibinfo {author}
  {\bibfnamefont {M.}~\bibnamefont {Brink}}, \bibinfo {author} {\bibfnamefont
  {A.}~\bibnamefont {Cross}}, \bibinfo {author} {\bibfnamefont {J.~M.}\
  \bibnamefont {Chow}},\ and\ \bibinfo {author} {\bibfnamefont {J.~M.}\
  \bibnamefont {Gambetta}},\ }\bibfield  {title} {\bibinfo {title}
  {Demonstration of weight-four parity measurements in the surface code
  architecture},\ }\href {https://doi.org/10.1103/PhysRevLett.117.210505}
  {\bibfield  {journal} {\bibinfo  {journal} {Phys. Rev. Lett.}\ }\textbf
  {\bibinfo {volume} {117}},\ \bibinfo {pages} {210505} (\bibinfo {year}
  {2016})}\BibitemShut {NoStop}%
\bibitem [{\citenamefont {Takita}\ \emph {et~al.}(2017)\citenamefont {Takita},
  \citenamefont {Cross}, \citenamefont {C\'orcoles}, \citenamefont {Chow},\
  and\ \citenamefont {Gambetta}}]{Takita17}%
  \BibitemOpen
  \bibfield  {author} {\bibinfo {author} {\bibfnamefont {M.}~\bibnamefont
  {Takita}}, \bibinfo {author} {\bibfnamefont {A.~W.}\ \bibnamefont {Cross}},
  \bibinfo {author} {\bibfnamefont {A.~D.}\ \bibnamefont {C\'orcoles}},
  \bibinfo {author} {\bibfnamefont {J.~M.}\ \bibnamefont {Chow}},\ and\
  \bibinfo {author} {\bibfnamefont {J.~M.}\ \bibnamefont {Gambetta}},\
  }\bibfield  {title} {\bibinfo {title} {Experimental demonstration of
  fault-tolerant state preparation with superconducting qubits},\ }\href
  {https://journals.aps.org/prl/abstract/10.1103/PhysRevLett.119.180501}
  {\bibfield  {journal} {\bibinfo  {journal} {Phys. Rev. Lett.}\ }\textbf
  {\bibinfo {volume} {119}},\ \bibinfo {pages} {180501} (\bibinfo {year}
  {2017})}\BibitemShut {NoStop}%
\bibitem [{\citenamefont {Harper}\ and\ \citenamefont
  {Flammia}(2019)}]{Harper19}%
  \BibitemOpen
  \bibfield  {author} {\bibinfo {author} {\bibfnamefont {R.}~\bibnamefont
  {Harper}}\ and\ \bibinfo {author} {\bibfnamefont {S.~T.}\ \bibnamefont
  {Flammia}},\ }\bibfield  {title} {\bibinfo {title} {Fault-tolerant logical
  gates in the ibm quantum experience},\ }\href
  {https://doi.org/10.1103/PhysRevLett.122.080504} {\bibfield  {journal}
  {\bibinfo  {journal} {Phys. Rev. Lett.}\ }\textbf {\bibinfo {volume} {122}},\
  \bibinfo {pages} {080504} (\bibinfo {year} {2019})}\BibitemShut {NoStop}%
\bibitem [{\citenamefont {Andersen}\ \emph {et~al.}(2019)\citenamefont
  {Andersen}, \citenamefont {Remm}, \citenamefont {Lazar}, \citenamefont
  {Krinner}, \citenamefont {Heinsoo}, \citenamefont {Besse}, \citenamefont
  {Gabureac}, \citenamefont {Wallraff},\ and\ \citenamefont
  {Eichler}}]{Kraglund19}%
  \BibitemOpen
  \bibfield  {author} {\bibinfo {author} {\bibfnamefont {C.~K.}\ \bibnamefont
  {Andersen}}, \bibinfo {author} {\bibfnamefont {A.}~\bibnamefont {Remm}},
  \bibinfo {author} {\bibfnamefont {S.}~\bibnamefont {Lazar}}, \bibinfo
  {author} {\bibfnamefont {S.}~\bibnamefont {Krinner}}, \bibinfo {author}
  {\bibfnamefont {J.}~\bibnamefont {Heinsoo}}, \bibinfo {author} {\bibfnamefont
  {J.-C.}\ \bibnamefont {Besse}}, \bibinfo {author} {\bibfnamefont
  {M.}~\bibnamefont {Gabureac}}, \bibinfo {author} {\bibfnamefont
  {A.}~\bibnamefont {Wallraff}},\ and\ \bibinfo {author} {\bibfnamefont
  {C.}~\bibnamefont {Eichler}},\ }\bibfield  {title} {\bibinfo {title}
  {{Entanglement stabilization using ancilla-based parity detection and
  real-time feedback in superconducting circuits}},\ }\href
  {https://doi.org/10.1038/s41534-019-0185-4} {\bibfield  {journal} {\bibinfo
  {journal} {npj Quantum Information}\ }\textbf {\bibinfo {volume} {5}},\
  \bibinfo {pages} {69} (\bibinfo {year} {2019})}\BibitemShut {NoStop}%
\bibitem [{\citenamefont {Ofek}\ \emph {et~al.}(2016)\citenamefont {Ofek},
  \citenamefont {Petrenko}, \citenamefont {Heeres}, \citenamefont {Reinhold},
  \citenamefont {Leghtas}, \citenamefont {Vlastakis}, \citenamefont {Liu},
  \citenamefont {Frunzio}, \citenamefont {Girvin}, \citenamefont {Jiang},
  \citenamefont {Mirrahimi}, \citenamefont {Devoret},\ and\ \citenamefont
  {Schoelkopf}}]{Ofek16}%
  \BibitemOpen
  \bibfield  {author} {\bibinfo {author} {\bibfnamefont {N.}~\bibnamefont
  {Ofek}}, \bibinfo {author} {\bibfnamefont {A.}~\bibnamefont {Petrenko}},
  \bibinfo {author} {\bibfnamefont {R.}~\bibnamefont {Heeres}}, \bibinfo
  {author} {\bibfnamefont {P.}~\bibnamefont {Reinhold}}, \bibinfo {author}
  {\bibfnamefont {Z.}~\bibnamefont {Leghtas}}, \bibinfo {author} {\bibfnamefont
  {B.}~\bibnamefont {Vlastakis}}, \bibinfo {author} {\bibfnamefont
  {Y.}~\bibnamefont {Liu}}, \bibinfo {author} {\bibfnamefont {L.}~\bibnamefont
  {Frunzio}}, \bibinfo {author} {\bibfnamefont {S.~M.}\ \bibnamefont {Girvin}},
  \bibinfo {author} {\bibfnamefont {L.}~\bibnamefont {Jiang}}, \bibinfo
  {author} {\bibfnamefont {M.}~\bibnamefont {Mirrahimi}}, \bibinfo {author}
  {\bibfnamefont {M.~H.}\ \bibnamefont {Devoret}},\ and\ \bibinfo {author}
  {\bibfnamefont {R.~J.}\ \bibnamefont {Schoelkopf}},\ }\bibfield  {title}
  {\bibinfo {title} {Extending the lifetime of a quantum bit with error
  correction in superconducting circuits},\ }\href
  {http://www.nature.com/nature/journal/v536/n7617/abs/nature18949.html}
  {\bibfield  {journal} {\bibinfo  {journal} {Nature}\ }\textbf {\bibinfo
  {volume} {536}},\ \bibinfo {pages} {441} (\bibinfo {year}
  {2016})}\BibitemShut {NoStop}%
\bibitem [{\citenamefont {Hu}\ \emph {et~al.}(2019)\citenamefont {Hu},
  \citenamefont {Ma}, \citenamefont {Cai}, \citenamefont {Mu}, \citenamefont
  {Xu}, \citenamefont {Wang}, \citenamefont {Wu}, \citenamefont {Wang},
  \citenamefont {Song}, \citenamefont {Zou}, \citenamefont {Girvin},
  \citenamefont {Duan},\ and\ \citenamefont {Sun}}]{Hu19}%
  \BibitemOpen
  \bibfield  {author} {\bibinfo {author} {\bibfnamefont {L.}~\bibnamefont
  {Hu}}, \bibinfo {author} {\bibfnamefont {Y.}~\bibnamefont {Ma}}, \bibinfo
  {author} {\bibfnamefont {W.}~\bibnamefont {Cai}}, \bibinfo {author}
  {\bibfnamefont {X.}~\bibnamefont {Mu}}, \bibinfo {author} {\bibfnamefont
  {Y.}~\bibnamefont {Xu}}, \bibinfo {author} {\bibfnamefont {W.}~\bibnamefont
  {Wang}}, \bibinfo {author} {\bibfnamefont {Y.}~\bibnamefont {Wu}}, \bibinfo
  {author} {\bibfnamefont {H.}~\bibnamefont {Wang}}, \bibinfo {author}
  {\bibfnamefont {Y.~P.}\ \bibnamefont {Song}}, \bibinfo {author}
  {\bibfnamefont {C.-L.}\ \bibnamefont {Zou}}, \bibinfo {author} {\bibfnamefont
  {S.~M.}\ \bibnamefont {Girvin}}, \bibinfo {author} {\bibfnamefont {L.-M.}\
  \bibnamefont {Duan}},\ and\ \bibinfo {author} {\bibfnamefont
  {L.}~\bibnamefont {Sun}},\ }\bibfield  {title} {\bibinfo {title} {{Quantum
  error correction and universal gate set operation on a binomial bosonic
  logical qubit}},\ }\href {https://doi.org/10.1038/s41567-018-0414-3}
  {\bibfield  {journal} {\bibinfo  {journal} {Nat.\ Phys.}\ }\textbf {\bibinfo
  {volume} {15}},\ \bibinfo {pages} {503} (\bibinfo {year} {2019})}\BibitemShut
  {NoStop}%
\bibitem [{\citenamefont {Negnevitsky}\ \emph {et~al.}(2018)\citenamefont
  {Negnevitsky}, \citenamefont {Marinelli}, \citenamefont {Mehta},
  \citenamefont {Lo}, \citenamefont {Fl{\"{u}}hmann},\ and\ \citenamefont
  {Home}}]{Negnevitsky18}%
  \BibitemOpen
  \bibfield  {author} {\bibinfo {author} {\bibfnamefont {V.}~\bibnamefont
  {Negnevitsky}}, \bibinfo {author} {\bibfnamefont {M.}~\bibnamefont
  {Marinelli}}, \bibinfo {author} {\bibfnamefont {K.~K.}\ \bibnamefont
  {Mehta}}, \bibinfo {author} {\bibfnamefont {H.-Y.}\ \bibnamefont {Lo}},
  \bibinfo {author} {\bibfnamefont {C.}~\bibnamefont {Fl{\"{u}}hmann}},\ and\
  \bibinfo {author} {\bibfnamefont {J.~P.}\ \bibnamefont {Home}},\ }\bibfield
  {title} {\bibinfo {title} {{Repeated multi-qubit readout and feedback with a
  mixed-species trapped-ion register}},\ }\href
  {https://doi.org/10.1038/s41586-018-0668-z} {\bibfield  {journal} {\bibinfo
  {journal} {Nature}\ }\textbf {\bibinfo {volume} {563}},\ \bibinfo {pages}
  {527} (\bibinfo {year} {2018})}\BibitemShut {NoStop}%
\bibitem [{\citenamefont {Knill}(2005)}]{Knill05}%
  \BibitemOpen
  \bibfield  {author} {\bibinfo {author} {\bibfnamefont {E.}~\bibnamefont
  {Knill}},\ }\bibfield  {title} {\bibinfo {title} {Quantum computing with
  realistically noisy devices},\ }\href {https://doi.org/10.1038/nature03350}
  {\bibfield  {journal} {\bibinfo  {journal} {Nature}\ }\textbf {\bibinfo
  {volume} {434}},\ \bibinfo {pages} {39} (\bibinfo {year} {2005})}\BibitemShut
  {NoStop}%
\bibitem [{\citenamefont {Saira}\ \emph {et~al.}(2014)\citenamefont {Saira},
  \citenamefont {Groen}, \citenamefont {Cramer}, \citenamefont {Meretska},
  \citenamefont {de~Lange},\ and\ \citenamefont {DiCarlo}}]{Saira14}%
  \BibitemOpen
  \bibfield  {author} {\bibinfo {author} {\bibfnamefont {O.-P.}\ \bibnamefont
  {Saira}}, \bibinfo {author} {\bibfnamefont {J.~P.}\ \bibnamefont {Groen}},
  \bibinfo {author} {\bibfnamefont {J.}~\bibnamefont {Cramer}}, \bibinfo
  {author} {\bibfnamefont {M.}~\bibnamefont {Meretska}}, \bibinfo {author}
  {\bibfnamefont {G.}~\bibnamefont {de~Lange}},\ and\ \bibinfo {author}
  {\bibfnamefont {L.}~\bibnamefont {DiCarlo}},\ }\bibfield  {title} {\bibinfo
  {title} {Entanglement genesis by ancilla-based parity measurement in 2{D}
  circuit {QED}},\ }\href
  {https://journals.aps.org/prl/abstract/10.1103/PhysRevLett.112.070502}
  {\bibfield  {journal} {\bibinfo  {journal} {Phys. Rev. Lett.}\ }\textbf
  {\bibinfo {volume} {112}},\ \bibinfo {pages} {070502} (\bibinfo {year}
  {2014})}\BibitemShut {NoStop}%
\bibitem [{Sup()}]{Suppmaterial}%
  \BibitemOpen
  \href@noop {} {}\bibinfo {note} {See supplementary materials.}\BibitemShut
  {Stop}%
\bibitem [{\citenamefont {McClure}\ \emph {et~al.}(2016)\citenamefont
  {McClure}, \citenamefont {Paik}, \citenamefont {Bishop}, \citenamefont
  {Steffen}, \citenamefont {Chow},\ and\ \citenamefont {Gambetta}}]{McClure16}%
  \BibitemOpen
  \bibfield  {author} {\bibinfo {author} {\bibfnamefont {D.~T.}\ \bibnamefont
  {McClure}}, \bibinfo {author} {\bibfnamefont {H.}~\bibnamefont {Paik}},
  \bibinfo {author} {\bibfnamefont {L.~S.}\ \bibnamefont {Bishop}}, \bibinfo
  {author} {\bibfnamefont {M.}~\bibnamefont {Steffen}}, \bibinfo {author}
  {\bibfnamefont {J.~M.}\ \bibnamefont {Chow}},\ and\ \bibinfo {author}
  {\bibfnamefont {J.~M.}\ \bibnamefont {Gambetta}},\ }\bibfield  {title}
  {\bibinfo {title} {Rapid driven reset of a qubit readout resonator},\ }\href
  {https://journals.aps.org/prapplied/abstract/10.1103/PhysRevApplied.5.011001}
  {\bibfield  {journal} {\bibinfo  {journal} {Phys. Rev. Appl.}\ }\textbf
  {\bibinfo {volume} {5}},\ \bibinfo {pages} {011001} (\bibinfo {year}
  {2016})}\BibitemShut {NoStop}%
\bibitem [{\citenamefont {O'Brien}\ \emph {et~al.}(2017)\citenamefont
  {O'Brien}, \citenamefont {Tarasinski},\ and\ \citenamefont
  {DiCarlo}}]{oBrien17}%
  \BibitemOpen
  \bibfield  {author} {\bibinfo {author} {\bibfnamefont {T.}~\bibnamefont
  {O'Brien}}, \bibinfo {author} {\bibfnamefont {B.}~\bibnamefont
  {Tarasinski}},\ and\ \bibinfo {author} {\bibfnamefont {L.}~\bibnamefont
  {DiCarlo}},\ }\bibfield  {title} {\bibinfo {title} {Density-matrix simulation
  of small surface codes under current and projected experimental noise},\
  }\href {https://www.nature.com/articles/s41534-017-0039-x} {\bibfield
  {journal} {\bibinfo  {journal} {npj Quantum Inf.}\ }\textbf {\bibinfo
  {volume} {3}},\ \bibinfo {pages} {39} (\bibinfo {year} {2017})}\BibitemShut
  {NoStop}%
\bibitem [{\citenamefont {Baum}\ and\ \citenamefont {Petrie}(1966)}]{Baum66}%
  \BibitemOpen
  \bibfield  {author} {\bibinfo {author} {\bibfnamefont {L.}~\bibnamefont
  {Baum}}\ and\ \bibinfo {author} {\bibfnamefont {T.}~\bibnamefont {Petrie}},\
  }\bibfield  {title} {\bibinfo {title} {Statistical inference for
  probabilistic functions of finite state markov chains},\ }\href
  {https://projecteuclid.org/euclid.aoms/1177699147} {\bibfield  {journal}
  {\bibinfo  {journal} {Ann. Math. Stat.}\ }\textbf {\bibinfo {volume} {37}},\
  \bibinfo {pages} {1554} (\bibinfo {year} {1966})}\BibitemShut {NoStop}%
\bibitem [{\citenamefont {{Akaike}}(1974)}]{Akaike74}%
  \BibitemOpen
  \bibfield  {author} {\bibinfo {author} {\bibfnamefont {H.}~\bibnamefont
  {{Akaike}}},\ }\bibfield  {title} {\bibinfo {title} {A new look at the
  statistical model identification},\ }\href
  {https://doi.org/10.1109/TAC.1974.1100705} {\bibfield  {journal} {\bibinfo
  {journal} {IEEE Transactions on Automatic Control}\ }\textbf {\bibinfo
  {volume} {19}},\ \bibinfo {pages} {716} (\bibinfo {year} {1974})}\BibitemShut
  {NoStop}%
\bibitem [{\citenamefont {Green}\ and\ \citenamefont {Swets}(1966)}]{Green66}%
  \BibitemOpen
  \bibfield  {author} {\bibinfo {author} {\bibfnamefont {D.~M.}\ \bibnamefont
  {Green}}\ and\ \bibinfo {author} {\bibfnamefont {J.~A.}\ \bibnamefont
  {Swets}},\ }\href@noop {} {\emph {\bibinfo {title} {Signal Detection Theory
  and Psychophysics}}}\ (\bibinfo  {publisher} {John Wiley \& Sons},\ \bibinfo
  {address} {New York},\ \bibinfo {year} {1966})\BibitemShut {NoStop}%
\bibitem [{\citenamefont {Magnard}\ \emph {et~al.}(2018)\citenamefont
  {Magnard}, \citenamefont {Kurpiers}, \citenamefont {Royer}, \citenamefont
  {Walter}, \citenamefont {Besse}, \citenamefont {Gasparinetti}, \citenamefont
  {Pechal}, \citenamefont {Heinsoo}, \citenamefont {Storz}, \citenamefont
  {Blais},\ and\ \citenamefont {Wallraff}}]{Magnard18}%
  \BibitemOpen
  \bibfield  {author} {\bibinfo {author} {\bibfnamefont {P.}~\bibnamefont
  {Magnard}}, \bibinfo {author} {\bibfnamefont {P.}~\bibnamefont {Kurpiers}},
  \bibinfo {author} {\bibfnamefont {B.}~\bibnamefont {Royer}}, \bibinfo
  {author} {\bibfnamefont {T.}~\bibnamefont {Walter}}, \bibinfo {author}
  {\bibfnamefont {J.-C.}\ \bibnamefont {Besse}}, \bibinfo {author}
  {\bibfnamefont {S.}~\bibnamefont {Gasparinetti}}, \bibinfo {author}
  {\bibfnamefont {M.}~\bibnamefont {Pechal}}, \bibinfo {author} {\bibfnamefont
  {J.}~\bibnamefont {Heinsoo}}, \bibinfo {author} {\bibfnamefont
  {S.}~\bibnamefont {Storz}}, \bibinfo {author} {\bibfnamefont
  {A.}~\bibnamefont {Blais}},\ and\ \bibinfo {author} {\bibfnamefont
  {A.}~\bibnamefont {Wallraff}},\ }\bibfield  {title} {\bibinfo {title} {Fast
  and unconditional all-microwave reset of a superconducting qubit},\ }\href
  {https://doi.org/10.1103/PhysRevLett.121.060502} {\bibfield  {journal}
  {\bibinfo  {journal} {Phys. Rev. Lett.}\ }\textbf {\bibinfo {volume} {121}},\
  \bibinfo {pages} {060502} (\bibinfo {year} {2018})}\BibitemShut {NoStop}%
\bibitem [{\citenamefont {Tomita}\ and\ \citenamefont
  {Svore}(2014)}]{Tomita14}%
  \BibitemOpen
  \bibfield  {author} {\bibinfo {author} {\bibfnamefont {Y.}~\bibnamefont
  {Tomita}}\ and\ \bibinfo {author} {\bibfnamefont {K.~M.}\ \bibnamefont
  {Svore}},\ }\bibfield  {title} {\bibinfo {title} {Low-distance surface codes
  under realistic quantum noise},\ }\href
  {https://link.aps.org/doi/10.1103/PhysRevA.90.062320} {\bibfield  {journal}
  {\bibinfo  {journal} {Phys. Rev. A}\ }\textbf {\bibinfo {volume} {90}},\
  \bibinfo {pages} {062320} (\bibinfo {year} {2014})}\BibitemShut {NoStop}%
\bibitem [{\citenamefont {Versluis}\ \emph {et~al.}(2017)\citenamefont
  {Versluis}, \citenamefont {Poletto}, \citenamefont {Khammassi}, \citenamefont
  {Tarasinski}, \citenamefont {Haider}, \citenamefont {Michalak}, \citenamefont
  {Bruno}, \citenamefont {Bertels},\ and\ \citenamefont
  {DiCarlo}}]{Versluis17}%
  \BibitemOpen
  \bibfield  {author} {\bibinfo {author} {\bibfnamefont {R.}~\bibnamefont
  {Versluis}}, \bibinfo {author} {\bibfnamefont {S.}~\bibnamefont {Poletto}},
  \bibinfo {author} {\bibfnamefont {N.}~\bibnamefont {Khammassi}}, \bibinfo
  {author} {\bibfnamefont {B.}~\bibnamefont {Tarasinski}}, \bibinfo {author}
  {\bibfnamefont {N.}~\bibnamefont {Haider}}, \bibinfo {author} {\bibfnamefont
  {D.~J.}\ \bibnamefont {Michalak}}, \bibinfo {author} {\bibfnamefont
  {A.}~\bibnamefont {Bruno}}, \bibinfo {author} {\bibfnamefont
  {K.}~\bibnamefont {Bertels}},\ and\ \bibinfo {author} {\bibfnamefont
  {L.}~\bibnamefont {DiCarlo}},\ }\bibfield  {title} {\bibinfo {title}
  {Scalable quantum circuit and control for a superconducting surface code},\
  }\href {https://doi.org/10.1103/PhysRevApplied.8.034021} {\bibfield
  {journal} {\bibinfo  {journal} {Phys. Rev. Appl.}\ }\textbf {\bibinfo
  {volume} {8}},\ \bibinfo {pages} {034021} (\bibinfo {year}
  {2017})}\BibitemShut {NoStop}%
\bibitem [{\citenamefont {Di{C}arlo}\ \emph {et~al.}(2009)\citenamefont
  {Di{C}arlo}, \citenamefont {Chow}, \citenamefont {Gambetta}, \citenamefont
  {Bishop}, \citenamefont {Johnson}, \citenamefont {Schuster}, \citenamefont
  {Majer}, \citenamefont {Blais}, \citenamefont {Frunzio}, \citenamefont
  {Girvin},\ and\ \citenamefont {Schoelkopf}}]{DiCarlo09}%
  \BibitemOpen
  \bibfield  {author} {\bibinfo {author} {\bibfnamefont {L.}~\bibnamefont
  {Di{C}arlo}}, \bibinfo {author} {\bibfnamefont {J.~M.}\ \bibnamefont {Chow}},
  \bibinfo {author} {\bibfnamefont {J.~M.}\ \bibnamefont {Gambetta}}, \bibinfo
  {author} {\bibfnamefont {L.~S.}\ \bibnamefont {Bishop}}, \bibinfo {author}
  {\bibfnamefont {B.~R.}\ \bibnamefont {Johnson}}, \bibinfo {author}
  {\bibfnamefont {D.~I.}\ \bibnamefont {Schuster}}, \bibinfo {author}
  {\bibfnamefont {J.}~\bibnamefont {Majer}}, \bibinfo {author} {\bibfnamefont
  {A.}~\bibnamefont {Blais}}, \bibinfo {author} {\bibfnamefont
  {L.}~\bibnamefont {Frunzio}}, \bibinfo {author} {\bibfnamefont {S.~M.}\
  \bibnamefont {Girvin}},\ and\ \bibinfo {author} {\bibfnamefont {R.~J.}\
  \bibnamefont {Schoelkopf}},\ }\bibfield  {title} {\bibinfo {title}
  {Demonstration of two-qubit algorithms with a superconducting quantum
  processor},\ }\href
  {http://www.nature.com/nature/journal/v460/n7252/abs/nature08121.html}
  {\bibfield  {journal} {\bibinfo  {journal} {Nature}\ }\textbf {\bibinfo
  {volume} {460}},\ \bibinfo {pages} {240} (\bibinfo {year}
  {2009})}\BibitemShut {NoStop}%
\bibitem [{\citenamefont {Jones}\ \emph {et~al.}(2001)\citenamefont {Jones},
  \citenamefont {Oliphant}, \citenamefont {Peterson} \emph {et~al.}}]{Scipy01}%
  \BibitemOpen
  \bibfield  {author} {\bibinfo {author} {\bibfnamefont {E.}~\bibnamefont
  {Jones}}, \bibinfo {author} {\bibfnamefont {T.}~\bibnamefont {Oliphant}},
  \bibinfo {author} {\bibfnamefont {P.}~\bibnamefont {Peterson}}, \emph
  {et~al.},\ }\href {http://www.scipy.org/} {\bibinfo {title} {{SciPy}: Open
  source scientific tools for {Python}}} (\bibinfo {year} {2001}),\ \bibinfo
  {note} {[Online; accessed 2016-06-14]}\BibitemShut {NoStop}%
\bibitem [{\citenamefont {Newville}\ \emph {et~al.}(2014)\citenamefont
  {Newville}, \citenamefont {Stensitzki}, \citenamefont {Allen},\ and\
  \citenamefont {Ingargiola}}]{LMFIT14}%
  \BibitemOpen
  \bibfield  {author} {\bibinfo {author} {\bibfnamefont {M.}~\bibnamefont
  {Newville}}, \bibinfo {author} {\bibfnamefont {T.}~\bibnamefont
  {Stensitzki}}, \bibinfo {author} {\bibfnamefont {D.~B.}\ \bibnamefont
  {Allen}},\ and\ \bibinfo {author} {\bibfnamefont {A.}~\bibnamefont
  {Ingargiola}},\ }\href {https://doi.org/10.5281/zenodo.11813} {\bibinfo
  {title} {{LMFIT: Non-Linear Least-Square Minimization and Curve-Fitting for
  Python¶}}} (\bibinfo {year} {2014})\BibitemShut {NoStop}%
\bibitem [{\citenamefont {Fu}\ \emph {et~al.}(2019)\citenamefont {Fu},
  \citenamefont {Riesebos}, \citenamefont {Rol}, \citenamefont {van Straten},
  \citenamefont {van Someren}, \citenamefont {Khammassi}, \citenamefont
  {Ashraf}, \citenamefont {Vermeulen}, \citenamefont {Newsum}, \citenamefont
  {Loh}, \citenamefont {de~Sterke}, \citenamefont {Vlothuizen}, \citenamefont
  {Schouten}, \citenamefont {Almudever}, \citenamefont {DiCarlo},\ and\
  \citenamefont {Bertels}}]{Fu19}%
  \BibitemOpen
  \bibfield  {author} {\bibinfo {author} {\bibfnamefont {X.}~\bibnamefont
  {Fu}}, \bibinfo {author} {\bibfnamefont {L.}~\bibnamefont {Riesebos}},
  \bibinfo {author} {\bibfnamefont {M.~A.}\ \bibnamefont {Rol}}, \bibinfo
  {author} {\bibfnamefont {J.}~\bibnamefont {van Straten}}, \bibinfo {author}
  {\bibfnamefont {J.}~\bibnamefont {van Someren}}, \bibinfo {author}
  {\bibfnamefont {N.}~\bibnamefont {Khammassi}}, \bibinfo {author}
  {\bibfnamefont {I.}~\bibnamefont {Ashraf}}, \bibinfo {author} {\bibfnamefont
  {R.~F.~L.}\ \bibnamefont {Vermeulen}}, \bibinfo {author} {\bibfnamefont
  {V.}~\bibnamefont {Newsum}}, \bibinfo {author} {\bibfnamefont {K.~K.~L.}\
  \bibnamefont {Loh}}, \bibinfo {author} {\bibfnamefont {J.~C.}\ \bibnamefont
  {de~Sterke}}, \bibinfo {author} {\bibfnamefont {W.~J.}\ \bibnamefont
  {Vlothuizen}}, \bibinfo {author} {\bibfnamefont {R.~N.}\ \bibnamefont
  {Schouten}}, \bibinfo {author} {\bibfnamefont {C.~G.}\ \bibnamefont
  {Almudever}}, \bibinfo {author} {\bibfnamefont {L.}~\bibnamefont {DiCarlo}},\
  and\ \bibinfo {author} {\bibfnamefont {K.}~\bibnamefont {Bertels}},\
  }\bibfield  {title} {\bibinfo {title} {{eQASM}: An executable quantum
  instruction set architecture},\ }in\ \href {https://arxiv.org/abs/1808.02449}
  {\emph {\bibinfo {booktitle} {Proceedings of 25th IEEE International
  Symposium on High-Performance Computer Architecture (HPCA)}}}\ (\bibinfo
  {organization} {IEEE},\ \bibinfo {address} {New York},\ \bibinfo {year}
  {2019})\ pp.\ \bibinfo {pages} {224--237}\BibitemShut {NoStop}%
\bibitem [{\citenamefont {Macklin}\ \emph {et~al.}(2015)\citenamefont
  {Macklin}, \citenamefont {O{\textquoteright}Brien}, \citenamefont {Hover},
  \citenamefont {Schwartz}, \citenamefont {Bolkhovsky}, \citenamefont {Zhang},
  \citenamefont {Oliver},\ and\ \citenamefont {Siddiqi}}]{Macklin15}%
  \BibitemOpen
  \bibfield  {author} {\bibinfo {author} {\bibfnamefont {C.}~\bibnamefont
  {Macklin}}, \bibinfo {author} {\bibfnamefont {K.}~\bibnamefont
  {O{\textquoteright}Brien}}, \bibinfo {author} {\bibfnamefont
  {D.}~\bibnamefont {Hover}}, \bibinfo {author} {\bibfnamefont {M.~E.}\
  \bibnamefont {Schwartz}}, \bibinfo {author} {\bibfnamefont {V.}~\bibnamefont
  {Bolkhovsky}}, \bibinfo {author} {\bibfnamefont {X.}~\bibnamefont {Zhang}},
  \bibinfo {author} {\bibfnamefont {W.~D.}\ \bibnamefont {Oliver}},\ and\
  \bibinfo {author} {\bibfnamefont {I.}~\bibnamefont {Siddiqi}},\ }\bibfield
  {title} {\bibinfo {title} {A near{\textendash}quantum-limited josephson
  traveling-wave parametric amplifier},\ }\href
  {http://www.sciencemag.org/cgi/doi/10.1126/science.aaa8525} {\bibfield
  {journal} {\bibinfo  {journal} {Science}\ }\textbf {\bibinfo {volume}
  {350}},\ \bibinfo {pages} {307} (\bibinfo {year} {2015})}\BibitemShut
  {NoStop}%
\bibitem [{\citenamefont {Bultink}\ \emph {et~al.}(2018)\citenamefont
  {Bultink}, \citenamefont {Tarasinski}, \citenamefont {Haandbaek},
  \citenamefont {Poletto}, \citenamefont {Haider}, \citenamefont {Michalak},
  \citenamefont {Bruno},\ and\ \citenamefont {DiCarlo}}]{Bultink18}%
  \BibitemOpen
  \bibfield  {author} {\bibinfo {author} {\bibfnamefont {C.~C.}\ \bibnamefont
  {Bultink}}, \bibinfo {author} {\bibfnamefont {B.}~\bibnamefont {Tarasinski}},
  \bibinfo {author} {\bibfnamefont {N.}~\bibnamefont {Haandbaek}}, \bibinfo
  {author} {\bibfnamefont {S.}~\bibnamefont {Poletto}}, \bibinfo {author}
  {\bibfnamefont {N.}~\bibnamefont {Haider}}, \bibinfo {author} {\bibfnamefont
  {D.~J.}\ \bibnamefont {Michalak}}, \bibinfo {author} {\bibfnamefont
  {A.}~\bibnamefont {Bruno}},\ and\ \bibinfo {author} {\bibfnamefont
  {L.}~\bibnamefont {DiCarlo}},\ }\bibfield  {title} {\bibinfo {title} {General
  method for extracting the quantum efficiency of dispersive qubit readout in
  circuit qed},\ }\href {https://doi.org/10.1063/1.5015954} {\bibfield
  {journal} {\bibinfo  {journal} {Appl. Phys. Lett.}\ }\textbf {\bibinfo
  {volume} {112}},\ \bibinfo {pages} {092601} (\bibinfo {year}
  {2018})}\BibitemShut {NoStop}%
\bibitem [{\citenamefont {Magesan}\ \emph {et~al.}(2012)\citenamefont
  {Magesan}, \citenamefont {Gambetta}, \citenamefont {Johnson}, \citenamefont
  {Ryan}, \citenamefont {Chow}, \citenamefont {Merkel}, \citenamefont
  {da~Silva}, \citenamefont {Keefe}, \citenamefont {Rothwell}, \citenamefont
  {Ohki}, \citenamefont {Ketchen},\ and\ \citenamefont {Steffen}}]{Magesan12b}%
  \BibitemOpen
  \bibfield  {author} {\bibinfo {author} {\bibfnamefont {E.}~\bibnamefont
  {Magesan}}, \bibinfo {author} {\bibfnamefont {J.~M.}\ \bibnamefont
  {Gambetta}}, \bibinfo {author} {\bibfnamefont {B.~R.}\ \bibnamefont
  {Johnson}}, \bibinfo {author} {\bibfnamefont {C.~A.}\ \bibnamefont {Ryan}},
  \bibinfo {author} {\bibfnamefont {J.~M.}\ \bibnamefont {Chow}}, \bibinfo
  {author} {\bibfnamefont {S.~T.}\ \bibnamefont {Merkel}}, \bibinfo {author}
  {\bibfnamefont {M.~P.}\ \bibnamefont {da~Silva}}, \bibinfo {author}
  {\bibfnamefont {G.~A.}\ \bibnamefont {Keefe}}, \bibinfo {author}
  {\bibfnamefont {M.~B.}\ \bibnamefont {Rothwell}}, \bibinfo {author}
  {\bibfnamefont {T.~A.}\ \bibnamefont {Ohki}}, \bibinfo {author}
  {\bibfnamefont {M.~B.}\ \bibnamefont {Ketchen}},\ and\ \bibinfo {author}
  {\bibfnamefont {M.}~\bibnamefont {Steffen}},\ }\bibfield  {title} {\bibinfo
  {title} {Efficient measurement of quantum gate error by interleaved
  randomized benchmarking},\ }\href
  {https://journals.aps.org/prl/abstract/10.1103/PhysRevLett.109.080505}
  {\bibfield  {journal} {\bibinfo  {journal} {Phys. Rev. Lett.}\ }\textbf
  {\bibinfo {volume} {109}},\ \bibinfo {pages} {080505} (\bibinfo {year}
  {2012})}\BibitemShut {NoStop}%
\bibitem [{\citenamefont {Ghosh}\ \emph {et~al.}(2013)\citenamefont {Ghosh},
  \citenamefont {Fowler}, \citenamefont {Martinis},\ and\ \citenamefont
  {Geller}}]{Ghosh13_B}%
  \BibitemOpen
  \bibfield  {author} {\bibinfo {author} {\bibfnamefont {J.}~\bibnamefont
  {Ghosh}}, \bibinfo {author} {\bibfnamefont {A.~G.}\ \bibnamefont {Fowler}},
  \bibinfo {author} {\bibfnamefont {J.~M.}\ \bibnamefont {Martinis}},\ and\
  \bibinfo {author} {\bibfnamefont {M.~R.}\ \bibnamefont {Geller}},\ }\bibfield
   {title} {\bibinfo {title} {Understanding the effects of leakage in
  superconducting quantum-error-detection circuits},\ }\href
  {https://doi.org/10.1103/PhysRevA.88.062329} {\bibfield  {journal} {\bibinfo
  {journal} {Phys. Rev. A}\ }\textbf {\bibinfo {volume} {88}},\ \bibinfo
  {pages} {062329} (\bibinfo {year} {2013})}\BibitemShut {NoStop}%
\bibitem [{\citenamefont {Waintal}(2019)}]{Waintal19}%
  \BibitemOpen
  \bibfield  {author} {\bibinfo {author} {\bibfnamefont {X.}~\bibnamefont
  {Waintal}},\ }\bibfield  {title} {\bibinfo {title} {What determines the
  ultimate precision of a quantum computer},\ }\href
  {https://doi.org/10.1103/PhysRevA.99.042318} {\bibfield  {journal} {\bibinfo
  {journal} {Phys. Rev. A}\ }\textbf {\bibinfo {volume} {99}},\ \bibinfo
  {pages} {042318} (\bibinfo {year} {2019})}\BibitemShut {NoStop}%
\bibitem [{qua()}]{quantumsim_website}%
  \BibitemOpen
  \href@noop {} {}\bibinfo {note} {The quantumsim package can be found at
  https://quantumsim.gitlab.io/}\BibitemShut {NoStop}%
\end{thebibliography}
%


\renewcommand{\theequation}{S\arabic{equation}}
\renewcommand{\thefigure}{S\arabic{figure}}
\renewcommand{\thetable}{S\arabic{table}}
\setcounter{figure}{0}
\setcounter{equation}{0}
\setcounter{table}{0}{}


\section*{Supplementary Materials for: Protecting quantum entanglement from leakage and qubit errors via repetitive parity measurements}
\
\section{Materials and methods}

\begin{figure*}[ht!]   
\centering     
\includegraphics{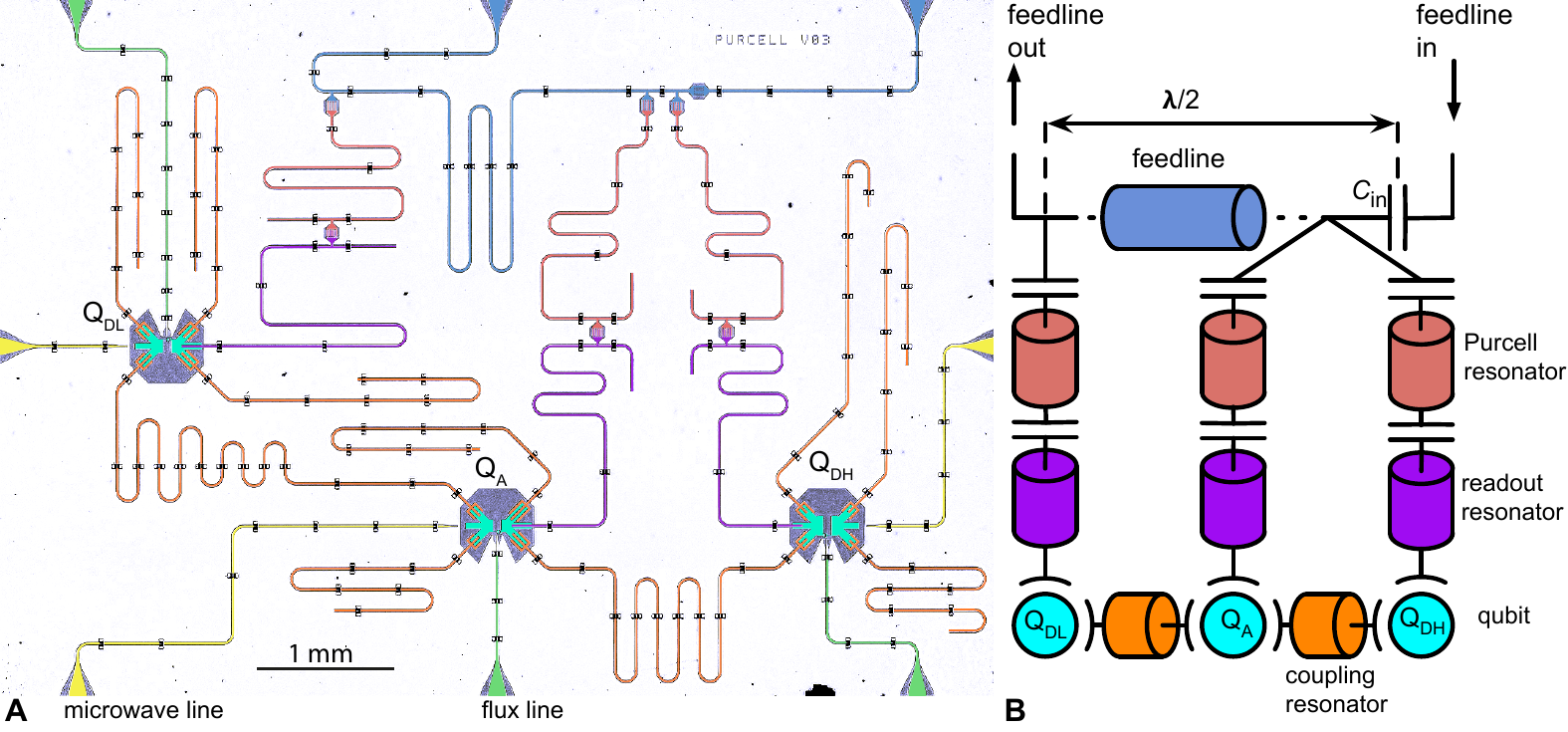}
\caption{\label{fig:device}   
False-colored photograph (\textbf{A}) and simplified circuit diagram (\textbf{B}) of the quantum processor with corresponding colors. 
} \end{figure*}

\subsection{Setup}
\begin{figure*}[!]   
\centering     
\includegraphics{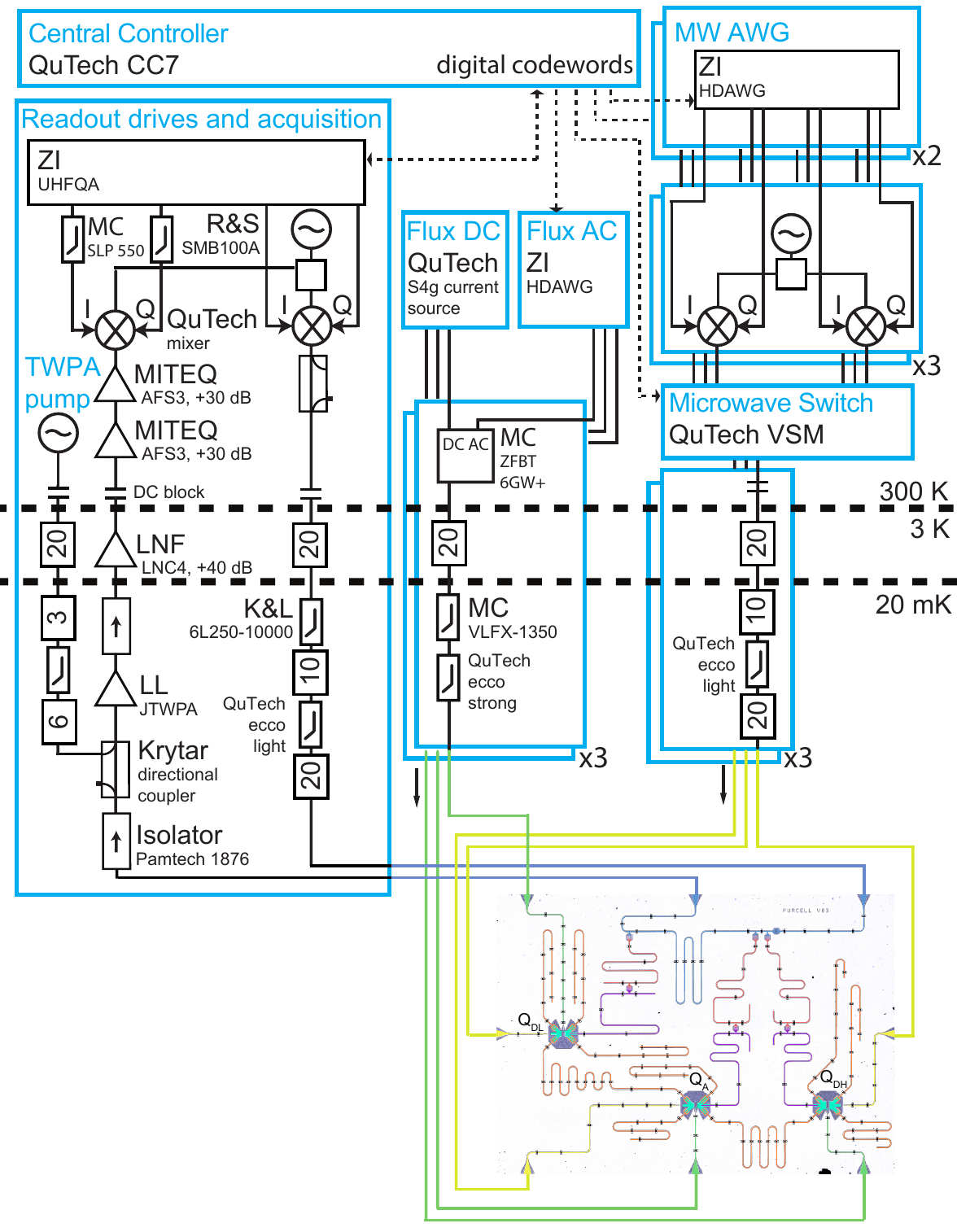}
\caption{\label{fig:setup}    
	Complete wiring diagram of electronic components inside and outside the 
	$^3$He/$^4$He dilution refrigerator (Leiden Cryogenics CF-CS81). 
} \end{figure*}

A full wiring diagram of the setup is provided in (\cref{fig:setup}). All operations are controlled by a fully digital device, the central controller (CC7), which takes as input a binary in an executable quantum instruction set architecture [eQASM~\cite{Fu19}], and outputs digital codeword triggers based on the execution result of these instructions. 
These digital codeword triggers are issued every 20~ns to arbitrary waveform generators (AWGs) for single-qubit gates and two-qubit gates, a vector switch matrix (VSM) for single-qubit gate routing and a readout module (AWG and acquisition) for frequency-multiplexed readout. 
Single-qubit gate generation, readout pulse generation and readout signal integration are performed by single-sideband mixing.
The measurement signal is amplified with a JTWPA~\cite{Macklin15} at the front end of the amplification chain. Following Ref.~\cite{Bultink18}, we extract an overall measurement efficiency $\eta=48\pm1.0\%$ by comparing the integrated signal-to-noise ratio of single-shot readout to the integrated measurement-induced dephasing.

\subsection{Cross-measurement-induced dephasing of data qubits}
\begin{figure}[ht!]   
\centering     
\includegraphics{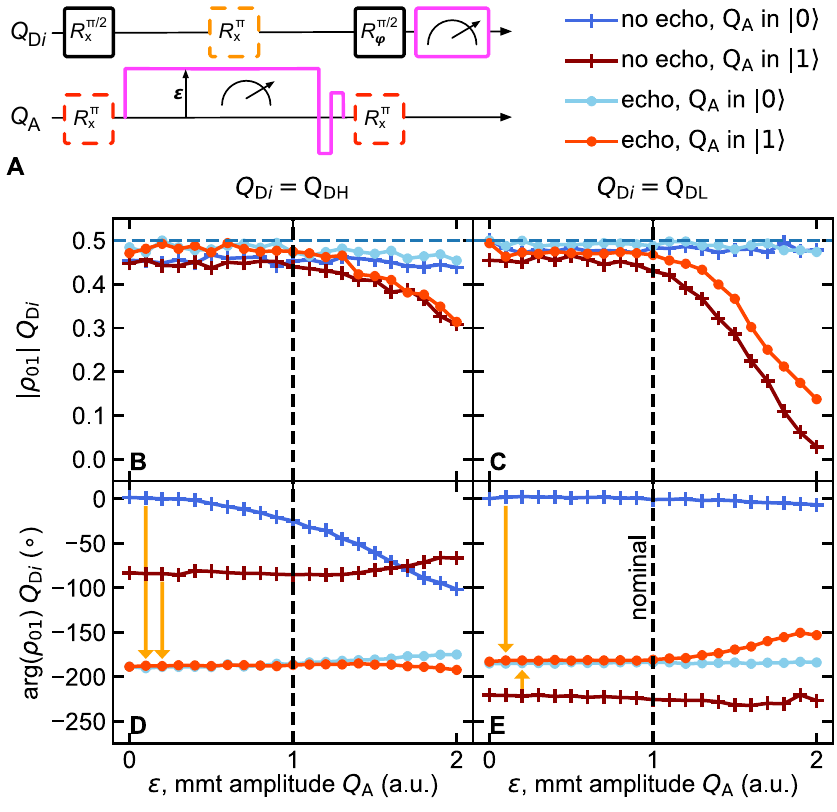}
\caption{\label{fig:dephasing}   
	Study of data-qubit coherence and phase accrual during ancilla measurement. 
	(\textbf{A})~Quantum circuit to extract data-qubit coherence and phase with or without echo pulse (orange) and with or without excitation in the ancilla. 
	(\textbf{B} and \textbf{C})~Data-qubit coherence as a function of ancilla measurement amplitude. 
	(\textbf{D} and \textbf{E})~Data-qubit phase as a function of ancilla measurement amplitude.
} \end{figure}

During ancilla measurement, data-qubit coherence is susceptible to intrinsic decoherence, phase shifts via residual ZZ interactions and cross-measurement-induced dephasing~\cite{Saira14,Heinsoo18}.
For the single-data-qubit subspace we investigate the different contributions experimentally and assess the benefit of an echo pulse on the data qubits halfway through the ancilla measurement. 
We study this by including the ancilla measurement (with amplitude $\varepsilon$) in a Ramsey-type sequence  (\cref{fig:dephasing}A). 
By varying the azimuthal phase of the second $\pi/2$ pulse, we obtain Ramsey fringes from which we extract the coherence $|\rho_{\mathrm{01}}|$ and phase $arg\left(\rho_{\mathrm{01}}\right)$. 
Several features of these curves explain the need for the echo pulse on the data qubits.
Firstly, at $\varepsilon=0$, the echo pulse improves data-qubit coherence (for both ancilla states) by reducing the effect of low-frequency noise (\cref{fig:dephasing}, B and C). This is confirmed by individual Ramsey and echo experiments. Secondly, the echo pulse almost perfectly cancels ancilla-state dependent phase shifts due to residual ZZ interactions (\cref{fig:dephasing}, D and E). 
When gradually turning on the ancilla measurement towards the nominal value $\varepsilon=1$, we furthermore observe that: thirdly, the echo pulse almost perfectly cancels the measurement-induced Stark shift (\cref{fig:dephasing}, D and E).
When increasing the measurement amplitude beyond the operation amplitude (indicated by the vertical dashed lines), we see rapid non-Gaussian decay of data-qubit coherence. 
We attribute this to measurement-induced relaxation of the ancilla: via the ZZ interaction, this can lead to probabilistic phase shifts on the data qubit. 
This effect is stronger for $\QDL$ than for $\QDH$ due to its higher residual interaction with $\QA$ (\cref{tab:device_parameters}).

\begin{table*}[!h]
   \begin{tabular}{|l|cccccc|}
      \hline
      \hline
      \multicolumn{1}{c}{} &~~~~  &~~~~ &~~~~  &~~~~ &~~~~ &~~~~ \\[\dimexpr-\normalbaselineskip-\arrayrulewidth]
      \textbf{Gate and Coherence Parameters} & \multicolumn{2}{c|}{$\bm{\QDL}$}& \multicolumn{2}{c|}{$\bm{\QA}$}& \multicolumn{2}{c|}{$\bm{\QDH}$}\\
      \hline 
      \hline
      {operating qubit frequency, $\omega_{\mathrm{op}}/2\pi$ (GHz)} & \multicolumn{2}{c|}{$5.02$}& \multicolumn{2}{c|}{$5.79$}& \multicolumn{2}{c|}{$6.88^{\dagger}$}\\
      \hline
      {max. qubit frequency, $\omega_{\mathrm{max}}/2\pi$ (GHz)} & \multicolumn{2}{c|}{$5.02$}& \multicolumn{2}{c|}{$5.79$}& \multicolumn{2}{c|}{$6.91$}\\
      \hline
      {anharmonicity, $\alpha/2\pi$ (MHz)} & \multicolumn{2}{c|}{$-306$}& \multicolumn{2}{c|}{$-308$}& \multicolumn{2}{c|}{$-331$}\\
      \hline
      {coherence time (at $\omega_{\mathrm{op}}/2\pi$), $T_2^{\mathrm{echo}}$ ($\mu$s)} & \multicolumn{2}{c|}{$29.6\pm2.7$}& \multicolumn{2}{c|}{$21.7\pm1.4$}& \multicolumn{2}{c|}{$14.7\pm0.9$}\\
      \hline
      {relaxation time (at $\omega_{\mathrm{op}}/2\pi$) $T_1$  ($\mu$s)} & \multicolumn{2}{c|}{$25.3\pm1.2$}& \multicolumn{2}{c|}{$17.0\pm0.6$}& \multicolumn{2}{c|}{$25.6\pm1.2$}\\
      \hline
      {Ramsey dephasing time (at $\omega_{\mathrm{op}}/2\pi$), $T_2^*$ ($\mu$s)} & \multicolumn{2}{c|}{$24.5\pm2.0$}& \multicolumn{2}{c|}{$14.6\pm1.2$}& \multicolumn{2}{c|}{$5.9\pm0.7$}\\
      \hline
      {average error per single qubit gate$^{\dagger\dagger\dagger\dagger}$, $e_{\mathrm{SQ}}$} (\%) & \multicolumn{2}{c|}{$0.08\pm0.02$}& \multicolumn{2}{c|}{$0.14\pm0.016$}& \multicolumn{2}{c|}{$0.21\pm0.06$}\\
      \hline
      {resonance exchange coupling, $J_1/2\pi$ (MHz)} & \multicolumn{3}{c|}{$17.2$}& \multicolumn{3}{c|}{14.3}\\
      \hline
      {bus resonator frequency, $\sim\omega_{\mathrm{bus}}/2\pi$ (GHz)} & \multicolumn{3}{c|}{$8.5$}& \multicolumn{3}{c|}{$8.5$}\\
      \hline
      {error per CZ$^{\dagger\dagger\dagger\dagger\dagger}$, $e_{\mathrm{CZ}}$ ($\%$)} & \multicolumn{3}{c|}{$1.4\pm0.6$}& \multicolumn{3}{c|}{$0.9\pm0.16$}\\
      \hline
      {leakage per CZ$^{\dagger\dagger\dagger\dagger\dagger}$, $L_{1}$ ($\%$)} & \multicolumn{3}{c|}{$0.27\pm0.12$}& \multicolumn{3}{c|}{$0.15\pm0.07$}\\
      \hline
      {ZZ coupling (at $\omega_{\mathrm{op}}/2\pi$), $\zeta_{\mathrm{ZZ}}/2\pi$ (MHz)} & \multicolumn{3}{c|}{$0.95$}& \multicolumn{3}{c|}{$0.33$}\\
      \hline
      \hline
      \textbf{Measurement Parameters} & \multicolumn{2}{c|}{$\bm{\QDL}$}& \multicolumn{2}{c|}{$\bm{\QA}$}& \multicolumn{2}{c|}{$\bm{\QDH}$}\\
      \hline 
      \hline
      {readout pulse frequency, $\omega_{\mathrm{ro}}/2\pi$} (GHz)& \multicolumn{2}{c|}{$7.225$}& \multicolumn{2}{c|}{$7.420$}& \multicolumn{2}{c|}{$7.838$}\\
      \hline
      {readout resonator frequency, $\omega_{\mathrm{ro}}/2\pi$} (GHz)& \multicolumn{2}{c|}{$7.275$}& \multicolumn{2}{c|}{$7.385$}& \multicolumn{2}{c|}{$7.867$}\\
      \hline
      {Purcell resonator frequency, $\omega_{\mathrm{ro}}/2\pi$} (GHz)& \multicolumn{2}{c|}{$7.260$}& \multicolumn{2}{c|}{$7.405$}& \multicolumn{2}{c|}{$7.872$}\\
      \hline
      {qubit-RR coupling strength, $g_{\mathrm{01,RR}}/2\pi$} (MHz)& \multicolumn{2}{c|}{$202$}& \multicolumn{2}{c|}{$188$}& \multicolumn{2}{c|}{$135$}\\
      \hline
      {PF-RR coupling strength, $J_{\mathrm{RR,PF}}/2\pi$} (MHz)& \multicolumn{2}{c|}{$48$}& \multicolumn{2}{c|}{$30$}& \multicolumn{2}{c|}{$38$}\\
      \hline
      {dispersive shift qubit-RR, $\chi_{\mathrm{RR}}/\pi$} (MHz)& \multicolumn{2}{c|}{$-2.5$}& \multicolumn{2}{c|}{$-5.3$}& \multicolumn{2}{c|}{$-2.8^{\dagger\dagger}$}\\
      \hline
      {dispersive shift qubit-PF, $\chi_{\mathrm{PF}}/\pi$} (MHz)& \multicolumn{2}{c|}{$-1.5$}& \multicolumn{2}{c|}{$-4.7$}& \multicolumn{2}{c|}{$-2.8^{\dagger\dagger}$}\\
      \hline
      {critical photon number, $n_{\mathrm{crit}}$}& \multicolumn{2}{c|}{2.3}& \multicolumn{2}{c|}{2.7}& \multicolumn{2}{c|}{$2.4$}\\
      \hline
      {intra-resonator photon number RR, $n_{\mathrm{RR}}$}& \multicolumn{2}{c|}{}& \multicolumn{2}{c|}{$1.2$}& \multicolumn{2}{c|}{}\\
      \hline
      {quantum efficiency, $\eta$ ($\%$)}& \multicolumn{2}{c|}{}& \multicolumn{2}{c|}{$48\pm1.0$}& 
      \multicolumn{2}{c|}{}\\
      \hline
      {Average assignment error, $e_{\mathrm{a}}$ ($\%$)}& \multicolumn{2}{c|}{$9.0^{\dagger\dagger\dagger}$}& \multicolumn{2}{c|}{$1.0\pm0.1$}& \multicolumn{2}{c|}{$16^{\dagger\dagger\dagger}$}\\
      \hline
      {Measurement integration time, $\tau_{\mathrm{int}}$ (ns)}& \multicolumn{2}{c|}{$600$}& \multicolumn{2}{c|}{$600$}& \multicolumn{2}{c|}{$600$}\\
      \hline
    \end{tabular}
    \caption{
        \label{tab:device_parameters}
        Measured parameters of the three-transmon device. $^{\dagger}$ $\QDH$ is operated $30$~MHz below its maximum frequency to avoid spurious interaction with a spurious two-level system. $^{\dagger\dagger}$ The Purcell mode and readout resonator mode of $\QDH$ have near-perfect hybridization (with qubit at $\omega_{\mathrm{op}}/2\pi$), making them indistinguishable. $^{\dagger\dagger\dagger}$ Single-shot readout on the data qubits was not optimized. $^{\dagger\dagger\dagger\dagger}$ Single-qubit gates are characterized using Clifford randomized benchmarking~\cite{Magesan12b} $^{\dagger\dagger\dagger\dagger\dagger}$ Two-qubit gates are characterized using interleaved RB~\cite{Magesan12b,Barends14} with a leakage-extraction modification~\cite{Rol19a}.
}
\end{table*}

\clearpage

\section{Supplementary Text}
\subsection{Performance of the simple hidden Markov model}
In this section we detail the performance of the simple HMMs, as described in \cref{fig:leakage}A and \cref{sec:MM.HMM.twostate} of the main text.
In \cref{fig:simple_model_performance}, A and B, we plot a histogram of the computational likelihoods $\Lcomp$ of $10^5$ simulated and actual $\SZZ$ experiments as calculated with the simple HMMs $\HZZ-D^{(\mathrm{simple})}$ and $\HZZ-A^{(\mathrm{simple})}$.
This can be compared with \cref{fig:leakage}, B and C, of the main text.
We plot similar histograms for the interleaved $\SZZ$---$\SXX$ experiment in \cref{fig:simple_model_performance}, H and I.
We see reasonable agreement, but noticeably worse agreement than that in the detailed model.
This is underscored by the Akaike information criterion (\cref{eq:akaike} of the main text), which is significantly reduced compared to the more detailed HMMs:
\begin{eqnarray}
A(\HZZ-D)-A(\HZZ-D^{(\mathrm{simple})})\\
=4.5\times 10^5 \\
A(\HZZ-A)-A(\HZZ-A^{(\mathrm{simple})})\\
=5.9\times 10^6 \\
A(\HZZXX-D)-A(\HZZXX-D^{(\mathrm{simple})})\\
=1.5\times 10^5 \\
A(\HZZXX-A)-A(\HZZXX-A^{(\mathrm{simple})})\\
=1.6\times 10^6.
\end{eqnarray}
Indeed, in all cases the Akaike information criterion for the simple HMM is lower than that for the detailed HMM without leakage.
This makes complete sense, as even though the simple HMMs might capture leakage fairly well, the additional effects captured in the detailed HMMs are far more dominant in the measurement signals than that of leakage.
As such, the internal metrics, such as the ROC curves (\cref{fig:ROCZZXXsimple}) for the simplified model are significantly less trustworthy than those of the detailed model.
This exemplifies the need for external HMM verification, as achieved in the main text by testing the HMM in a leakage mitigation scheme.
We now repeat this verification procedure for the simple model.
We see that in the $\SZZ$ experiment the performance is significantly degraded; although the flat line in the $\ZZ$ curve is restored after about $8$ parity checks, it requires rejecting $47\%$ of the data, and is restored to a point $\sim 8\%$ below the performance of the detailed HMM.
By contrast, the simple HMM performs almost identically to the complex HMM in the interleaved $\SZZ$---$\SXX$ experiment, achieving Bell-state fidelities within $2\%$ whilst retaining the same amount of data.
As the signal from a large-scale QEC code is more similar to the latter experiment than the former (See \cref{sec:future}), this strongly suggests that the detailed modeling used in this text will not be needed in such experiments.

\subsection{Hidden Markov models for large-scale QEC}\label{sec:future}
The hidden Markov models used in this text provide an exciting prospect for the indirect detection of leakage on both data qubits and ancillas in a QEC code.
This is essential for accurate decoding of stabilizer measurements made during QEC.
Furthermore, this idea can be combined with proposals for leakage reduction~\cite{Aliferis07,Fowler13,Ghosh15,Suchara15} to target such efforts, reducing unnecessary overhead.
As leakage does not spread in superconducting qubits (to lowest order), and gives only local error signals~\cite{Ghosh15}, such a scheme would require a single HMM per (data and ancilla) qubit.
Each individual HMM needs only to process the local error syndrome, and as demonstrated in this work, completely independent HMMs may be used for the detection of nearby data-qubit and ancilla leakage.
This implies that the computational overhead of leakage detection via HMMs in a larger QEC code will grow only linearly with the system size.
Previous leakage reduction units are designed to act as the identity on the computational subspace (up to additional noise), so we do not require perfect discrimination between leaked and computational states.
However, optimizing this discrimination (and investigating threshold levels for the application of targeted leakage reduction) will boost the code performance.
Also, near-perfect discrimination could allow for the direct resetting of leaked data qubits~\cite{Magnard18}, which would completely destroy an error correcting code if not targeted.

On the other hand, for implementation on classical hardware within the sub-$1~\mu s$ QEC cycle time on superconducting qubits~\cite{oBrien17}, one may wish to strip back some of the optimization used in this work.
The minimal HMM that could be used in QEC for detection has only two states, leaked and unleaked (\cref{fig:leakage}A), and $2^{\nA}$ outputs, where $\nA$ is the number of neighboring ancilla on which a signature of leakage is detected.
(For the surface code, $\nA\leq 4$ in all situations.)
Such a simple model cannot perfectly deal with correlated errors, such as ancilla errors (which give multiple error signals separated in time).
However, this should only cause a slight reduction in the discrimination capability whenever such correlations remain local.
If the loss in accuracy is acceptable, one may store only $\pi_0^{(\mathrm{post})}$, and update it following a measurement $\MA[m]$ as
\begin{eqnarray}
    &\pi_0^{(\mathrm{prior})}[m]\\
    &=(A_{0,0}-A_{0,1})\pi_0^{(\mathrm{post})}[m-1]+A_{0,1},\\
    &\pi_0^{(\mathrm{post})}[m]\\
    &=\frac{\pi_0^{(\mathrm{prior})}[m]B_{\MA[m],0}}{B_{\MA[m],1}+\pi_0^{(\mathrm{prior})}(B_{\MA[m],0}-B_{\MA[m],1})},
\end{eqnarray}
which is trivial compared to the overhead for most QEC decoders.

A key question about the use of HMMs for leakage detection in future QEC experiments is whether leakage in larger codes is reliably detectable.
In previous theoretical work~\cite{Ghosh13_B}, data-qubit leakage in repetition codes has been sometimes hidden, a phenomenon known as `leakage paralysis' or `silent stabilizer'~\cite{Waintal19}.
This effect occurs when the relative phase $\varphi$ accumulated between the $|20\rangle$ and $|21\rangle$ states during a CZ gate is a multiple of $\pi$.
In the absence of additional error, an indirect measurement of the data qubit via an ancilla would return a result $\frac{\varphi}{\pi}$ mod $2$.
(By comparison, if $\varphi=\pi/2$, the ancilla would return measurements of $0$ or $1$ at random.)
This is then identical to the measurement of a data qubit in the $|\frac{\varphi}{\pi}\;\mathrm{mod}\;2\rangle$ state, and no discrimination between the two may be achieved.
However, in an $N$-qubit parity check $S$, the ancilla continues to accumulate phase from the other qubits, reducing this to an $N-1$-qubit effective parity check $S'$ (plus a well-defined, constant phase).
Such a parity check may no longer commute with other effective parity checks $R'$ that share the leaked qubit, even though we would require $[S,R]=0$ in stabilizer QEC.
This is demonstrated in our second experiment measuring both $\SZZ$ and $\SXX$ parity checks; though these commute when no data qubit is leaked, leakage reduces the checks to non-commuting $Z$ and $X$ measurements (of the unleaked data qubit).
(In the $\SZZ$ experiment, the leakage paralysis was broken by the echo pulse on the data qubits, which flips the effective stabilizer of a leaked qubit at each round.)
The repeated measurement of these non-commuting operators generates random results, similar to the case when $\varphi=\pi/2$.
To the best of our knowledge, in all fully fault-tolerant stabilizer QEC codes, the removal of a single data qubit breaks the commutativity of at least two neighboring stabilizers.
As such, data-qubit leakage will always be detectable in QEC experiments with superconducting circuits.

Beyond the proof-of-principle argument above, one might question whether the signal of leakage is improved or reduced when going from our prototype experiment to a larger QEC code, and when the underlying physical-qubit error rate is reduced.
Fortunately, we can expect an improvement in the HMM discrimination capability in both situations.
To see this, consider the example of a data qubit which is either leaked at round $1$ with probability $ \pleak$ or never leaks.
Let us further assume that in the absence of leakage, a number of neighboring ancillas $n_{\mathrm{A}}$ incur errors (where the parity check reports a flip) at a rate $p$, whereas in the presence of leakage these ancillas incur errors at a rate $0.5$.
(For example, in the bulk of the surface code, $n_{\mathrm{A}}=4$.)
The computational likelihood at round $m>0$ after seeing $e$ errors may be calculated as
\begin{equation}
    \Lcomp[m]=\frac{(1- \pleak)p^e(1-p)^{m \nA-e}}{(1- \pleak)p^e(1-p)^{m \nA-e}+ \pleak(0.5)^{m \nA}}.
\end{equation}
If the data qubit was leaked, $e\sim m \nA/2$, and the computational likelihood on average is approximately
\begin{equation}
    L_{\mathrm{comp}}[m]\sim \frac{1- \pleak}{ \pleak}\left(\frac{p^{ \nA/2}(1-p)^{ \nA/2}}{0.5^{ \nA}}\right)^m,
\end{equation}
which is of the form
\begin{eqnarray}
    L_{\mathrm{comp}}[m]=&Ae^{-\lambda m},\hspace{0.5cm} A=\frac{1- \pleak}{ \pleak},\\
     \lambda=&\log\left(2^{ \nA}p^{-\frac{ \nA}{2}}(1-p)^{-\frac{ \nA}{2}}\right).
\end{eqnarray}
We see that the signal of leakage ($L_{\mathrm{comp}}[m]\rightarrow 0$) switches on exponentially in time, with a rate proportional to $\log(p^{- \nA/2})$.
Any decrease in $p$ (from better qubits) or increases in $ \nA$ (from additional ancillas surrounding the leaked qubit in a QEC code) will serve to increase, and not decrease this rate.
The exponential decay constant is inversely proportional to the leakage rate (as this corresponds to an initial HMM skepticism towards unlikely leakage events).
However, as the likelihood 'switch' is exponential, a decrease in $ \pleak$ by even an order of magnitude should only increase the time before definite detection by a single step or so.
The above analysis is complicated in a real scenario, as single physical errors give correlated detection signals, and as leakage may occur at any time, and as leaked qubits may seep.
Correlations in the detection signals will serve to renormalize the switching time $\lambda$ (but not remove the generic feature of exponential onset).
Seepage causes individual leakage events to be finite (with some average lifetime $T_{\mathrm{seep}}$); an individual leakage event of length $\ll\lambda^{-1}$ will not be detectable by the HMM.
However, when the system returns to the computational subspace in such a short period of time, the leakage event may be treated as a `regular' error, and does not need complicated leakage-detection hardware for fault tolerance.
For example, a leakage event followed by immediate decay to $\ket{1}$ is indistinguishable from a direct transition to $\ket{1}$ for all practical purposes in QEC.

\section{Additional Figures}
\newpage

\begin{figure}[!]   
\centering     
\includegraphics{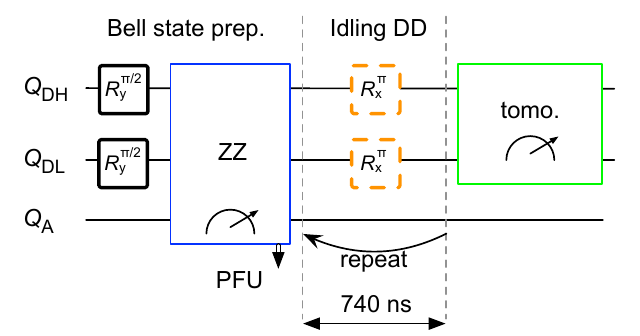}
\caption{\label{fig:idling}    
	Quantum circuit for Bell-state idling experiments under dynamical decoupling. 
} \end{figure}

\begin{figure}[!]   
\centering     
\includegraphics{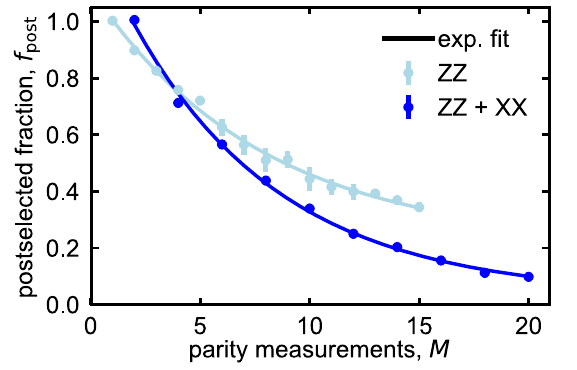}
\caption{\label{fig:fractions}    
	Postselected fractions for the 'no error' conditioning in \cref{fig:ZZs,fig:ZZXXs}.
} \end{figure}

\begin{figure}[ht!]   
\centering     
\includegraphics{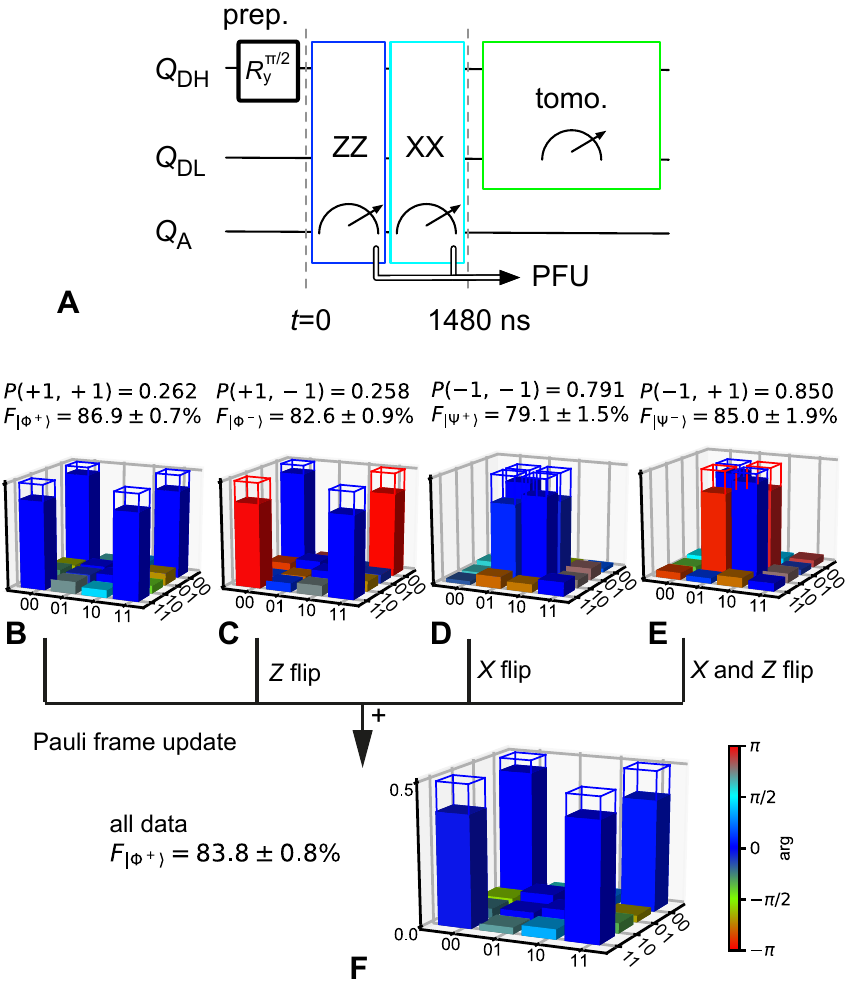}
\caption{\label{fig:ZZXX} Generating entanglement by sequential $\SZZ$ and $\SXX$ parity measurements and PFU. 
(\textbf{A})~Simplified quantum circuit for preparation, $\SZZ$ and $\SXX$ measurements, sequential data-qubit state tomography and PFU. 
(\textbf{B} to \textbf{E})~Manhattan-style plots of the reconstructed data-qubit density matrix conditioned on the ancilla measurement outcomes with occurrence and fidelity to the four expected Bell states. 
(\textbf{F})~We use the two-bit outcome of the parity checks to apply a PFU that transforms all runs ideally to $\ket{\Phi^+}$. 
Frames on the tomograms indicate the Bell states ideally produced.
} 
\end{figure}

\begin{figure}[ht!]   
\centering     
\includegraphics{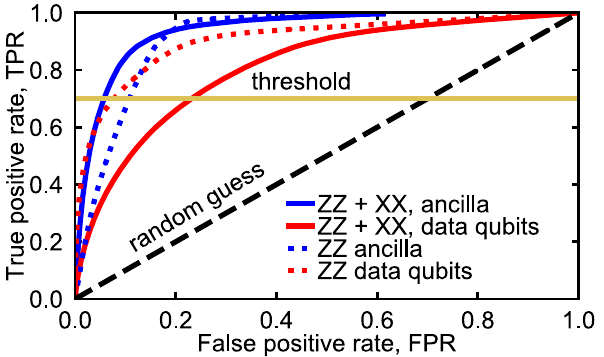}
\caption{\label{fig:ROCZZXX} Receiver operating characteristics (ROCs) for mitigation of data-qubit and ancilla leakage during interleaved $\SZZ$ and $\SXX$ checks. Data-qubit and ancilla leakage are each discerned via a dedicated HMM (full curves). For comparison, the ROCs for the HMMs for repeated $\SZZ$ checks only are also shown (dotted curves, same data as in \cref{fig:leakage}F).}
\end{figure}

\begin{figure*}[ht!]   
\centering     
\includegraphics{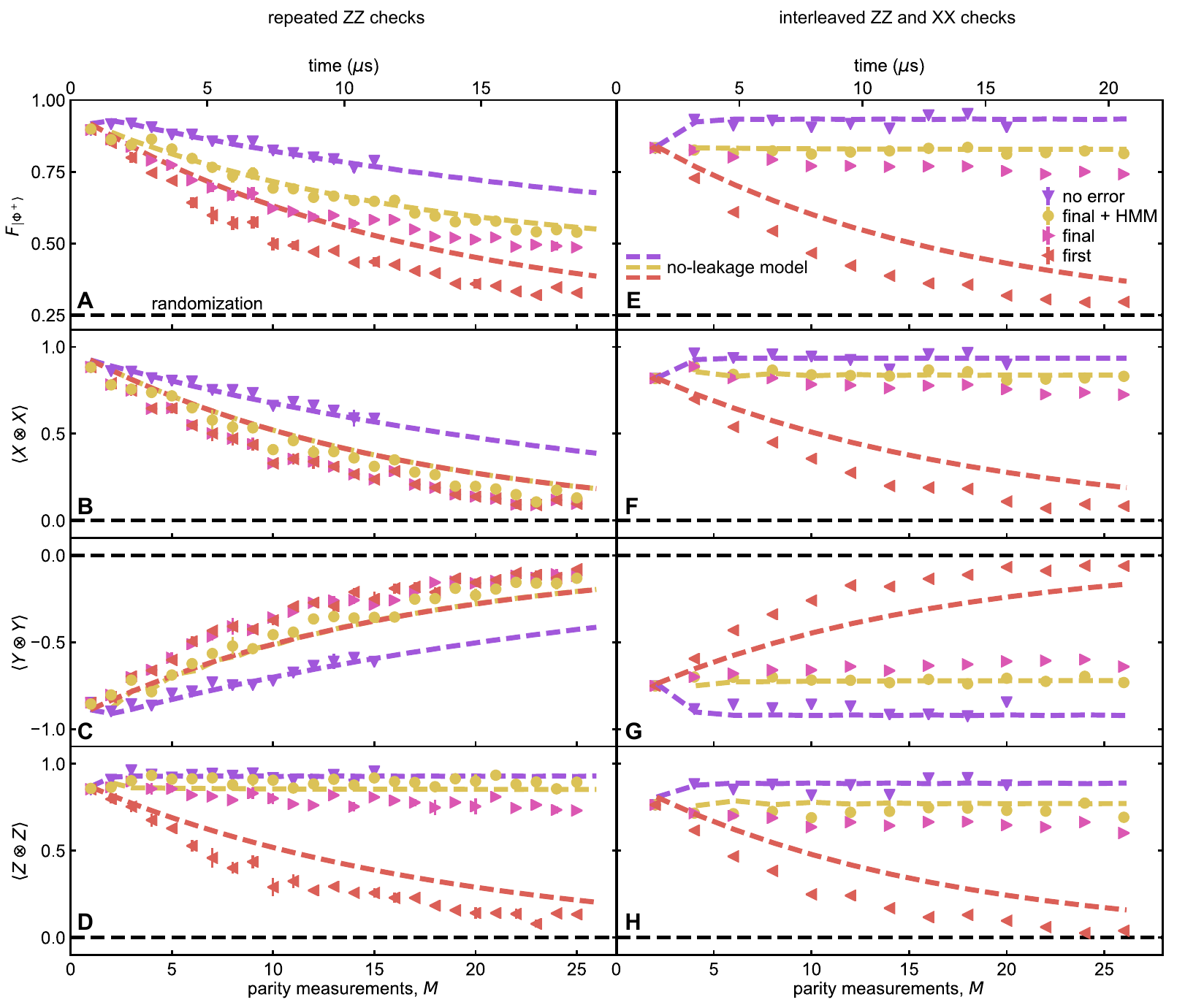}
\caption{\label{fig:no_leakage_model} 
Comparison of experimental data and no-leakage modeling of the repeated parity check experiments of \cref{fig:ZZs,fig:ZZXXs}. Simulations use the independently measured  $T_2^{\mathrm{echo}}$, $T_1$, $e_{\mathrm{a}}$, $e_{\mathrm{SQ}}$, $e_{\mathrm{CZ}}$) of \cref{tab:device_parameters}. This modeling uses two-level systems (no leakage) following Ref.~\cite{oBrien17}, which uses quantumsim~\cite{quantumsim_website}. As expected, the modeling is outperforming the experiment for `first' and `final' correction strategies as the modeling does not include leakage. It however shows an excellent matching for the `no error' conditioning (which rejects both qubit errors and leakage). The `final + HMM' is excellently matching the `final' modeling curve, confirming the leakage detection capability of the HMMs.
}
\end{figure*}

\begin{figure*}[ht!]   
\centering     
\includegraphics{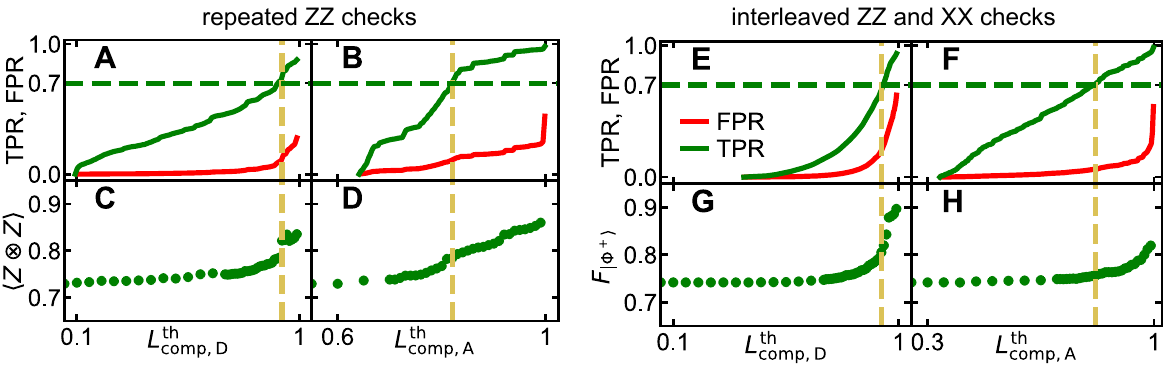}
\caption{\label{fig:L_tradeoff} 
Leakage mitigation for the repeated parity check experiments as a function of the chosen threshold.
(\textbf{A}) [(\textbf{B})] $\TPR$, $\FPR$ as a function of the chosen computational-space likelihood threshold for the repeated parity check experiments of  \cref{fig:ZZs,fig:leakage} for data-qubit leakage [ancilla leakage] at $M=25$. 
(\textbf{C}) [(\textbf{D})] The improvement  in repeated $\SZZ$ checks is expressed as the increase in $\ZZ$ for data-qubit leakage [ancilla leakage]. 
Horizontal dashed lines indicate the chosen threshold $\TPR=0.7$ (\cref{fig:leakage}, F and G) and vertical dashed lines indicate the accompanying computational-space likelihoods. 
(\textbf{E}-\textbf{H}) Similar plots for leakage rejection for interleaved $\SZZ$ and $\SXX$ checks (\cref{fig:ZZXXs}) at $M=26$. The protocol improvement is here expressed as an increase of $\FBell$.
}
\end{figure*}

\begin{figure*}[ht!]   
\centering     
\includegraphics{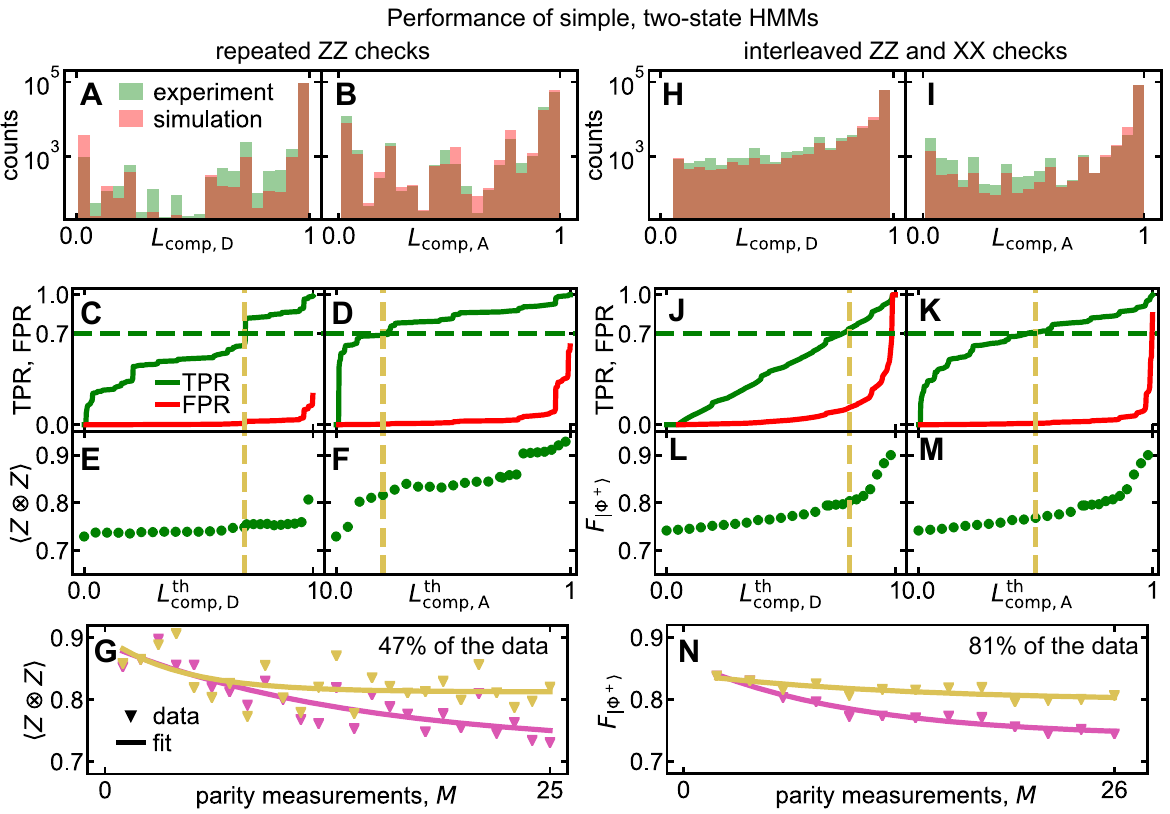}
\caption{\label{fig:simple_model_performance} 
Leakage mitigation for the simple, two-state HMMs for repeated parity check experiments as a function of the chosen threshold.
(\textbf{A}) [(\textbf{B})]~Histograms of $10^5$ $\vecs$ with $M=25$ for repeated $\SZZ$ checks (as in \cref{fig:leakage}D [\cref{fig:leakage}E]). 
HMM training suggests $3.6\%$ [$20\%$] total data-qubit [ancilla] leakage at $M=25$. 
(\textbf{C}) [(\textbf{D})] $\TPR$, $\FPR$ as a function of the chosen computational-space likelihood threshold for the repeated parity check experiments of \cref{fig:ZZs} for data-qubit leakage [ancilla leakage] at $M=25$. 
(\textbf{E}) [(\textbf{F})] The improvement in repeated $\SZZ$ checks is expressed as the increase in $\ZZ$ for data-qubit leakage [ancilla leakage]. 
Horizontal dashed lines indicate the chosen threshold $\TPR=0.7$ and vertical dashed lines indicate the accompanying computational-space likelihoods (as in \cref{fig:L_tradeoff}). 
(\textbf{G}) $\ZZ$ after $M$ $\SZZ$ checks and correction based on the `final' outcomes, without (same data as in \cref{fig:ZZs}D) and with leakage mitigation by postselection ($\TPR=0.7$).
(\textbf{H}-\textbf{M}) Similar plots for simple-HMM leakage rejection for interleaved $\SZZ$ and $\SXX$ checks (\cref{fig:ZZXXs}) at $M=26$. 
(\textbf{N}) $\FBell$ after $M$ interleaved checks and correction based on the `final' outcomes, without (same data as in \cref{fig:ZZXXs}) and with leakage mitigation by postselection ($\TPR=0.7$). The protocol improvement (L, M and N) is here expressed as an increase of $\FBell$.} 
\end{figure*}

\begin{figure}[ht!]   
\centering     
\includegraphics{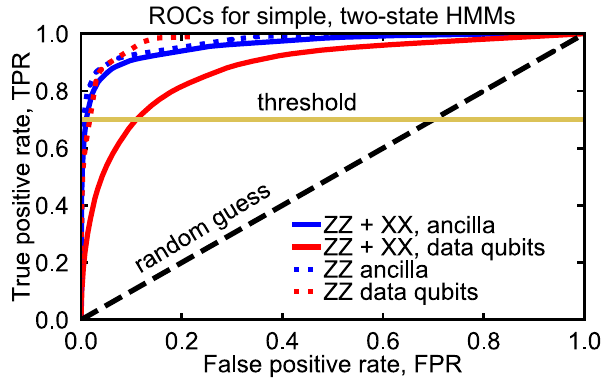}
\caption{\label{fig:ROCZZXXsimple} 
Receiver operating characteristics (ROCs) for leakage mitigation as in \cref{fig:ROCZZXX}, but using simple two-state HMMs. }
\end{figure}

\end{document}